# Revisiting relativistic magnetohydrodynamics from quantum electrodynamics

Masaru Hongo,[a,b] Koichi Hattori[c]

[a]*Department of Physics, University of Illinois, Chicago, IL 60607, USA*
[b]*RIKEN iTHEMS, RIKEN, Wako 351-0198, Japan*
[c]*Yukawa Institute for Theoretical Physics, Kyoto University, Kyoto 606-8502, Japan*

*E-mail:* masaru.hongo@riken.jp, koichi.hattori@yukawa.kyoto-u.ac.jp

ABSTRACT: We provide a statistical mechanical derivation of relativistic magnetohydrodynamics on the basis of $(3+1)$-dimensional quantum electrodynamics; the system endowed with a magnetic one-form symmetry. The conservation laws and constitutive relations are presented in a manifestly covariant way with respect to the general coordinate transformation. The method of the local Gibbs ensemble (or nonequilibrium statistical operator) combined with the path-integral formula for a thermodynamic functional enables us to obtain exact forms of constitutive relations. Applying the derivative expansion to exact formulas, we derive the first-order constitutive relations for nonlinear relativistic magnetohydrodynamics. Our results for the QED plasma preserving parity and charge-conjugation symmetries are equipped with two electrical resistivities and five (three bulk and two shear) viscosities. We also show that those transport coefficients satisfy the Onsager's reciprocal relation and a set of inequalities, indicating semi-positivity of the entropy production rate consistent with the local second law of thermodynamics.

## Contents

## 1  Introduction

Establishment of quantum electrodynamics (QED) is a milestone at the dawn of diverse developments in quantum field theories. The success of the covariant perturbative renormalization *a la* Tomonaga, Schwinger, Feynmann, and Dyson [1–8] (together with later developments in the renormalization-group method [9–14]) has deepened our fundamental theoretical backgrounds in modern physics. The further success of the perturbation theory for the electron (lepton) anomalous magnetic moment [15] has reached the state-of-art technology as the most accurate theoretical calculation presented by humankind [16–19]. Besides, QED has played a role of the prototype of quantum gauge theories: generalization to non-Abelian theories has led us to quantum chromodynamics (QCD) and electroweak theory—other basic building blocks of the standard model—that contain richer nonperturbative dynamics induced by, e.g., the asymptotic freedom and the Higgs mechanism.

In applications of QED to many-body systems, real-time dynamics of the plasma has been one of the central issues and still occupies one theoretical hot place. Such plasmas are composed of the electron and proton (and/or positron) in astrophysics [20, 21] and high-intensity laser physics [22, 23] and of the electron and hole in table-top materials in condensed matter physics. In the weak-coupling regime, real-time dynamics of plasmas may be described with a quasi-particle approximation, leading to the well-established kinetic theory, or the Boltzmann equation [24]. However, this is not always the case because plasmas in, e.g., a certain kind of materials has an effectively large electromagnetic coupling, which invalidates the weak-coupling approximation. A general formalism for describing the strong-coupling plasma has not been established thus far (See, e.g., Refs. [25–27] and references therein for reviews on strong-coupling phenomena beyond the quasi-particle approximations and condensed matter applications of the holographic principle). Therefore, it is temping to develop the hydrodynamic framework that works even in the strong-coupling regime when we focus on the low-energy (long-time/long-length) dynamics [28]. The presence of a dynamical electromagnetic field modifies the usual hydrodynamic equation to that of *magnetohydrodynamics* as the low-energy effective theory.

The conventional formulation of magnetohydrodynamics is simply to combine two famous equations, i.e., the Navier-Stokes equation for a fluid and the Maxwell equation for an electromagnetic field (See, e.g., Refs. [29–32] for representative textbooks). In the relativistic formulation, they are given as [33, 34]

$$\partial_\mu T_{\text{matt}}^{\mu\nu} = F^{\nu\mu} j_\mu \,, \tag{1.1a}$$

$$\partial_\mu F^{\mu\nu} = -j^\nu \,, \tag{1.1b}$$

where $T_{\text{matt}}^{\mu\nu}$ and $F^{\mu\nu}$ are the energy-momentum tensor of the matter (fluid) and the electromagnetic field strength tensor. The right-hand sides of those equations are the Joule-heat/Lorentz-force term and the electrical current $j^\nu$ which provide sources of the energy/momentum and the electric field, respectively. They are "non-conservation equations" and contain a gapped mode that dissipates in time, i.e., the electric field damped out by the Debye screening effect. Therefore, the hydrodynamic variables (or, conserved charge densities) followed from global symmetries of the system, are not appropriately identified in such a conventional formulation. An electric field should be, instead, induced by the dynamics of hydrodynamic variables, and thus, obey a constitutive relation. Another aspect of this issue is that energy-momentum tensors of the matter $T_{\text{matt}}^{\mu\nu}$ and the electromagnetism are introduced separately[1]. However, what the translational symmetry of the system tells us is the conservation of the *total* energy and momentum composed of a mixture of matter and electromagnetic fields, instead of the matter component alone. Those fundamental issues have been overlooked in the conventional formulation[2].

Recent progress in reformulation of magnetohydrodynamics has clarified those theoretical issues [35–39]. The vital point is to recognize the Bianchi identity as the conservation law that describes time-evolution of a magnetic flux density. Moreover, motivated by the generalized symmetry viewpoint, the presence of the Bianchi identity is tied to the so-called magnetic one-form symmetry in the $(3+1)$-dimensional QED [35–39]. One-form symmetry leads to the conservation of a one-dimensional object, i.e., the magnetic flux lines in the QED plasma and is classified as a particular case of higher-form symmetry [40]. As a consequence of this generalized viewpoint of global symmetry, one can regard the magnetic flux density as a canonical conserved-charge density. However, the electric flux is not endowed with such a role since the Maxwell equation (1.1b) has the electric current as the source term. Based on this observation, relativistic magnetohydrodynamics has been formulated as the low-energy effective theory for dynamics of the conserved charge densities associated with the Poincaré symmetry and magnetic one-form symmetry.

---

[1]Namely, eliminating the electric current $j^\mu$ in Eqs. (1.1a) and (1.1b) and using the Bianchi identity, one can render them in the conservation equation: $\partial_\mu (T_{\text{matt}}^{\mu\nu} + T_{\text{Maxwell}}^{\mu\nu}) = 0$, where the Maxwell stress tensor is given by $T_{\text{Maxwell}}^{\mu\nu} = F^{\mu\alpha} F^\nu_{\ \alpha} - \eta^{\mu\nu} F^{\alpha\beta} F_{\alpha\beta}/4$ with the Minkowski metric $\eta_{\mu\nu} = \text{diag}(-1, +1, +1, +1)$. Clearly, one finds a separation of the electromagnetic part from the matter part $T_{\text{matt}}^{\mu\nu}$.

[2]Nevertheless, the presence of those theoretical issues do *not* necessarily means that the conventional approach fails to describe magnetohydrodynamic behaviors. Eliminating the electric field, one can indeed extract magnetohydrodynamic equations as long as the separation between matter and electromagnetic parts of the energy-momentum tensor works as a good approximation.

Despite this conceptual development, most of the studies still crucially rely on the assumption for the local laws of thermodynamics in a macroscopic point of view, leaving a phenomenological flavor in the derivation of the hydrodynamic equations [35, 36, 38, 39] (see Ref. [36] for the QED plasma with the chiral anomaly). On the other hand, there has been a remarkable field-theoretical development that (partially) overcomes the phenomenological assumption [41, 42]. In this approach, the energy-momentum tensor is obtained from the variation of the hydrostatic partition function with respect to the background curved metric, which enables us to derive a *nondissipative* part of constitutive relations[3]. Formulation of magnetohydrodynamics with the hydrostatic method was carried out in Ref. [34], *albeit*, in the aforementioned conventional way of coupling the fluid to the electromagnetic field. Similarly, the recent attempt to derive relativistic magnetohydrodynamics from the Boltzmann-Vlasov equation [51, 52], which relies on the phenomenological collision terms, is not fully based on the global symmetry QED as in the present paper.

The purpose of this paper is to lift the phenomenological framework to a statistical mechanical level [53–61] and reconcile it with the underlying field-theoretical description of QED [33, 62–66]. Based on the symmetries and conservation laws inherent in the QED plasma, we identify a Gibbs-type density operator in describing the local equilibrium state with the dynamical magnetic field. Correct identification of the local Gibbs density operators enables us to apply the local Gibbs ensemble (or the so-called nonequilibrium statistical operator) method to establish the field-theoretical derivation of the relativistic magnetohydrodynamics at a nonlinear level[4]. We provide an exact expression for the constitutive relation, path-integral formula for local thermal equilibrium, and then perform a derivative expansion to derive the first-order dissipative corrections to the local equilibrium dynamics. In addition to the constitutive relations, we derive the Green-Kubo formulas [67–69], Onsager's reciprocal relation [70] and the inequalities for transport coefficients on the basis of the local Gibbs ensemble method. As a corollary of the inequalities, we also confirm semi-positivity of the local entropy production rate. The present paper completes the field-theoretical derivation of nonlinear relativistic dissipative magnetohydrodynamics in the exact form that systematically generates the derivative expansions of constitutive relations on an order-by-order basis. This development builds a solid foundation for the preceding phenomenological derivations in the literature.

---

[3]See also Refs. [43–50] for the description of the dissipative fluid with effective field theories based on the Schwinger-Keldysh formalism.

[4]By "nonlinear" we here mean that the derived magnetohydrodynamic equation is a deterministic nonlinear partial differential equation. Linearization on top of a static background leads to linearized magnetohydrodynamics, which contains propagating modes like the magnetosonic and Alfvén waves.

The organization of the paper is as follows: In Sec. 2, we briefly review the conservation laws for the QED plasma from its global symmetries equipped with background gauge fields. In Sec. 3, we present a basic strategy of the local Gibbs ensemble method to derive the magnetohydrodynamic equation. Section 4 is devoted to the exact results both for the local equilibrium and off-equilibrium parts. In Sec. 5, we perform a derivative expansion and obtain the first-order constitutive relations supplemented with the Green-Kubo formulas for transport coefficients. The Onsager's reciprocal relation and the inequalities are also explicitly shown there (The comparison of those results to the previous works is given in Appendix A.). In Sec. 6, we summarize the results and discuss prospects.

## 2 Symmetries of QED and conservation laws

We consider $(3+1)$-dimensional QED plasmas such as an electron-positron plasma. Its microscopic dynamics is governed by the following Lagrangian:

$$\mathcal{L}_{\text{QED}} = -\bar{\psi}\gamma^\mu(\partial_\mu + \mathrm{i}qA_\mu)\psi - m\bar{\psi}\psi - \frac{1}{4}\eta^{\mu\nu}\eta^{\alpha\beta}F_{\mu\alpha}F_{\nu\beta}, \qquad (2.1)$$

where $\psi$ denotes a charged fermionic field with a mass $m$ and electric charge $q$ ($q$ is negative for a negatively charged fermion). For simplicity, we assume that $\psi$ represents a single component Dirac field (with the gamma matrices $\gamma^\mu$), but generalization to, e.g., multicomponent cases with or without charged scalar fields is straightforward. We also introduced the U(1) gauge field $A_\mu$ with the field strengh tensor $F_{\mu\nu} \equiv \partial_\mu A_\nu - \partial_\nu A_\mu$. The QED coupling constant $e$ is included in the definition of charge $q$, and $\eta_{\mu\nu}$ is the Minkowski metric with the mostly-plus convention, $\eta_{\mu\nu} = \text{diag}(-1,+1,+1,+1)$. For brevity, we will use a collective notation for dynamical variables (fermion and photon fields) as $\varphi = \{\psi, \bar{\psi}, A_\mu\}$.

The goal of this paper is to derive the hydrodynamic equations on the basis of the QED Lagrangian (2.1). Since a set of hydrodynamic equations consists of conservation laws, the first step is to specify the corresponding continuous *global* symmetry of QED. One can immediately identify the Poincaré invariance, i.e., the spacetime translation symmetry and Lorentz symmetry, as such symmetries resulting in the conservation law for the energy-momentum and total angular momentum. As shown shortly, only the energy-momentum conservation serves as a dynamical equation, whereas the angular-momentum conservation turns out to be a constraint equation.

Besides, the QED Lagrangian (2.1) enjoys the so-called magnetic one-form symmetry (often called the generalized global symmetry). While our original dynamical variables $\varphi$ are inert under the magnetic one-form symmetry, it acts on a dual gauge field $\widetilde{A}_\mu$ defined by $\partial_\mu \widetilde{A}_\nu - \partial_\nu \widetilde{A}_\mu \equiv \frac{1}{2}\epsilon^{\mu\nu\rho\sigma}F_{\mu\nu}$ with a totally anti-symmetric tensor $\epsilon^{\mu\nu\rho\sigma}$ ($\epsilon^{0123} = 1$). The magnetic one-form symmetry generates a shift of the dual gauge field $\widetilde{A}_\mu \to \widetilde{A}_\mu + \theta_\mu$, and the corresponding conservation law is found to be

$$\partial_\mu \widetilde{F}^{\mu\nu} = 0. \tag{2.2}$$

This is nothing but the Bianchi identity. This equation means that one can regard the magnetic flux as a conserved quantity associated with the magnetic one-form symmetry. Notice that we distinguish the magnetic flux from a magnetic field. The latter will be introduced as a conjugate variable in the subsequent section. In contrast, the QED Lagrangian (2.1) is not invariant under the electric one-form symmetry, which generates a shift of the original gauge field $A_\mu$. The explicit breaking of the electric one-form symmetry is owing to the presence of a charged matter, so that the Maxwell equation with the electric current (1.1b) describes non-conservation of the electric flux[5]. One thus only needs to take into account dynamics of the energy-momentum density and magnetic flux density in magnetohydrodynamics.

Here is another remark: Despite its conserving property, the electric charge density does not serve as a hydrodynamic (or gapless) variable in the presence of *dynamical* electromagnetic fields. Thus, the U(1) gauge symmetry does not yield a gapless hydrodynamic mode due to its non-local nature. In other words, a finite electric charge density induces a gapped mode, i.e., an electric flux density. This clearly causes a conflict with the absence of an electric flux in equilibrium states without spontaneous symmetry breaking such as the charge density wave state. The local equilibrium is realized only after they are damped out[6], so that a net electric charge density does not appear in magnetohydrodynamics. Both an electric current (including an electric charge density) and electric flux are thus induced by dynamics of hydrodynamic variables as already mentioned below Eq. (1.1).

As widely known, global symmetries of the system can be tied to conservation laws in a systematic and useful way by introducing the corresponding background fields[7]. For this purpose, we introduce the background fields associated with the Poincaré and magnetic one-form symmetries, which results in the following gauged action:

$$\mathcal{S}_{\mathrm{QED}}[\varphi; j] = \int \mathrm{d}^4 x \sqrt{-g} \left[ -\frac{1}{2} \bar{\psi} \left( \gamma^a e_a{}^\mu \overrightarrow{D}_\mu - \overleftarrow{D}_\mu \gamma^a e_a{}^\mu \right) \psi - m \bar{\psi} \psi - \frac{1}{4} g^{\mu\nu} g^{\alpha\beta} F_{\mu\alpha} F_{\nu\beta} + \frac{1}{2} b_{\mu\nu} \widetilde{F}^{\mu\nu} \right]. \tag{2.3}$$

We summarize generalities of the gauged action (2.3) as follows. To gauge the spacetime symmetries, we introduced a vierbein field $e_\mu{}^a$ and its inverse $e_a{}^\mu$. Throughout this paper, we use the Greek letters ($\mu, \nu, \ldots = 0, 1, 2, 3$) for the curved spacetime indices, and the Latin letters ($a, b, \ldots = 0, 1, 2, 3$) for the local Lorentz indices. The vierbein specifies the local relation between the curved metric $g_{\mu\nu}(x)$ and the Minkowski metric $\eta_{ab} = \mathrm{diag}(-1, +1, +1, +1)$ as $g_{\mu\nu}(x) = e_\mu{}^a(x) e_\nu{}^b(x) \eta_{ab}$, so that its determinant gives a correct measure $\sqrt{-g} = \det e_\mu{}^a$ with $g \equiv \det g_{\mu\nu} < 0$.

---

[5]The free Maxwell theory in the absence of a charged matter also enjoys the electric one-form symmetry, and the electric flux is conserved.

[6]Intuitively, one may not put like-sign charges in a finite distance, which costs a finite Coulomb energy. Therefore, the QED plasma in an infinite volume system cannot reach a thermal (or even steady) state with a finite electric charge density coupled to the dynamical U(1) gauge field.

[7]Besides, it will serve as a solid basis to discuss transport phenomena taking place in the local thermal equilibrium as discussed in Sec. 4.1.

We also defined the covariant derivative of the Dirac field as

$$\overrightarrow{D}_\mu \psi \equiv \partial_\mu \psi + \mathrm{i}q A_\mu \psi - \frac{\mathrm{i}}{2}\omega_\mu{}^{ab}\Sigma_{ab}\psi \quad \text{and} \quad \bar{\psi}\overleftarrow{D}_\mu \equiv \partial_\mu \bar{\psi} - \mathrm{i}q\bar{\psi} A_\mu + \frac{\mathrm{i}}{2}\bar{\psi}\omega_\mu{}^{ab}\Sigma_{ab}, \quad (2.4)$$

where we introduced a spin connection $\omega_\mu{}^{ab}$ upon a requirement of the covariance under the local Lorentz transformation. Recalling the transformation property of the Dirac field, one can identify its representation matrix $\Sigma_{ab}$ as $\Sigma_{ab} = \mathrm{i}[\gamma_a, \gamma_b]/4$. Within the vierbein postulate, $D_\mu e_\nu{}^a = \nabla_\mu e_\nu{}^a + \omega_\mu{}^a{}_b e_\nu{}^b = 0$, and the torsinonless assumption, one can express the spin connection by the vierbein as

$$\omega_\mu{}^{ab} = \frac{1}{2} e^{a\nu} e^{b\rho}(C_{\nu\rho\mu} - C_{\rho\nu\mu} - C_{\mu\nu\rho}), \quad (2.5)$$

where we introduced the Ricci rotation coefficient $C_{\mu\nu\rho} \equiv e_\mu{}^c(\partial_\nu e_{\rho c} - \partial_\rho e_{\nu c})$. We also introduced a background two-form field $b_{\mu\nu}$ minimally coupled to the magnetic current $\widetilde{F}^{\mu\nu}$ defined by

$$\widetilde{F}^{\mu\nu} \equiv \frac{1}{2}\varepsilon^{\mu\nu\rho\sigma} F_{\rho\sigma}. \quad (2.6)$$

Here, we used a normalized totally anti-symmetric tensor $\varepsilon^{\mu\nu\rho\sigma} = \epsilon^{\mu\nu\rho\sigma}/\sqrt{-g}$ in the curved spacetime. Note that $\varepsilon^{\mu\nu\rho\sigma}$ (instead of $\epsilon^{\mu\nu\rho\sigma}$) behaves as a tensor under the general coordinate transformation. In the following, we will also use a collective notation to express a set of background fields as $j \equiv \{e_\mu{}^a, b_{\mu\nu}\}$.

Now, based on the gauged action (2.3), let us derive the covariant conservation laws in the presence of background fields. The crucial point here is that the gauged action (2.3) remains invariant ($\delta_\chi \mathcal{S}_\text{QED} = 0$) under the local transformation acting on both the dynamical and background fields:

$$\begin{cases} \delta_\chi \psi(x) = \xi^\mu \partial_\mu \psi - \dfrac{\mathrm{i}}{2}\alpha^{ab}\Sigma_{ab}\psi, \\ \delta_\chi A_\mu(x) = \xi^\nu \nabla_\nu A_\mu + A_\nu \nabla_\mu \xi^\nu, \\ \delta_\chi e_\mu{}^a(x) = \xi^\nu \nabla_\nu e_\mu{}^a + e_\nu{}^a \nabla_\mu \xi^\nu + \alpha^a{}_b e_\mu{}^b, \\ \delta_\chi b_{\mu\nu}(x) = \xi^\rho \nabla_\rho b_{\mu\nu} + b_{\rho\nu}\nabla_\mu \xi^\rho + b_{\mu\rho}\nabla_\nu \xi^\rho + \partial_\mu \theta_\nu - \partial_\nu \theta_\mu. \end{cases} \quad (2.7)$$

Here, we introduced a set of infinitesimal local transformation parameters $\chi \equiv \{\xi^\mu, \alpha^a{}_b, \theta_\mu\}$ for a general coordinate ($\xi^\mu$), local Lorentz ($\alpha^{ab} = -\alpha^{ba}$), and magnetic one-form gauge transformation ($\theta_\mu$), respectively. On the other hand, (re)introducing a set of gauge currents in terms of functional derivatives of the action (2.3) as

$$T^\mu{}_a(t, \boldsymbol{x}) \equiv \frac{1}{e}\frac{\delta \mathcal{S}_\text{QED}}{\delta e_\mu{}^a(t, \boldsymbol{x})} \quad \text{and} \quad J^{\mu\nu}(t, \boldsymbol{x}) \equiv \frac{2}{e}\frac{\delta \mathcal{S}_\text{QED}}{\delta b_{\mu\nu}(t, \boldsymbol{x})}, \quad (2.8)$$

we obtain another expression for the variation of the action (for notational simplicity, we hereafter express the magnetic current as $J^{\mu\nu} \equiv \widetilde{F}^{\mu\nu}$):

$$\delta_\chi \mathcal{S}_\text{QED} = \int \mathrm{d}^4 x \left[\frac{\delta \mathcal{S}_\text{QED}}{\delta e_\mu{}^a(x)}\delta_\chi e_\mu{}^a(x) + \frac{\delta \mathcal{S}_\text{QED}}{\delta b_{\mu\nu}(x)}\delta_\chi b_{\mu\nu}(x) + \frac{\delta \mathcal{S}_\text{QED}}{\delta \varphi(x)}\delta_\chi \varphi(x)\right]$$

$$= -\int \mathrm{d}^4 x \sqrt{-g}\Big[\xi^\nu \Big(\nabla_\mu T^\mu{}_\nu - \frac{1}{2}J^{\alpha\beta}(\nabla_\nu b_{\alpha\beta} + \nabla_\beta b_{\nu\alpha} + \nabla_\alpha b_{\beta\nu})-\frac{1}{2}(T_{ab}-T_{ba})\omega_\nu{}^{ab}\Big)$$

$$+ \frac{1}{2}(T_{ab} - T_{ba})\alpha^{ab} + \theta_\nu \nabla_\mu J^{\mu\nu} \Big] + \text{(surface terms)}. \tag{2.9}$$

To reach the second line, we performed an integration by parts and used the equation of motion $\delta \mathcal{S}_{\mathrm{QED}}/\delta\varphi = 0$. Since the action is invariant ($\delta_\chi \mathcal{S}_{\mathrm{QED}} = 0$) under the transformation with an arbitrary $\chi = \{\xi^\mu, \alpha^{ab}, \theta_\nu\}$, we find conservation laws of the energy-momentum tensor and magnetic flux

$$\nabla_\mu T^\mu{}_\nu = \frac{1}{2} J^{\alpha\beta} H_{\nu\alpha\beta}, \quad \nabla_\mu J^{\mu\nu} = 0 \quad \text{with} \quad H_{\nu\alpha\beta} \equiv \nabla_\nu b_{\alpha\beta} + \nabla_\beta b_{\nu\alpha} + \nabla_\alpha b_{\beta\nu}, \tag{2.10}$$

together with the following constraint on the energy-momentum tensor:

$$T^{ab} - T^{ba} = 0. \tag{2.11}$$

The source term $J^{\alpha\beta} H_{\nu\alpha\beta}$ appearing on the right-hand side of the energy-momentum equation (2.10) is an analogue of the Lorentz-force term induced by the background electromagnetic field.

We have thus identified the conservation laws (2.10) associated with global symmetries of the QED plasma. Those two equations govern the low-energy dynamics of the conserved charge densities, that is, the energy-momentum density and magnetic flux density. In the following sections, we will derive hydrodynamic equations based on operator versions of the conservation laws (2.10), which finally constitute magnetohydrodynamics.

## 3 Local Gibbs ensemble method for magnetohydrodynamics

In this section, we present our setup and oveview of the local Gibbs ensemble method in deriving relativistic magnetohydrodynamics. After introducing a crucial assumption on the initial density operator, we explain the difficulty of the problem and how the use of the local Gibbs ensemble method overcomes that.

### 3.1 Local Gibbs ensemble

Based on the preparation in the last section, we here describe the basic framework of the local Gibbs ensemble method [63–65] applied to our formulation of magnetohydrodynamics. Our starting point is operator versions of the conservation laws in Eq. (2.10):

$$\nabla_\mu \hat{T}^\mu{}_\nu = \frac{1}{2} \hat{J}^{\alpha\beta} H_{\nu\alpha\beta}, \quad \nabla_\mu \hat{J}^{\mu\nu} = 0, \tag{3.1}$$

where $H_{\nu\alpha\beta}$ remains a classical external field defined in Eq. (2.10). While these equations provide operator identities, which serves as a basis of hydrodynamics, we still need to specify a state (or density operator) to derive hydrodynamic equations. Therefore, we, here, make a crucial assumption that the system is in a *local* thermal equilibrium at the initial time $t_0$. One can describe a local thermal equilibrium state with the local Gibbs distribution

$$\hat{\rho}_{\mathrm{LG}}[t; \lambda_t] = \exp(-\hat{S}[t; \lambda_t]), \tag{3.2}$$

where we defined the entropy operator $\hat{S}[t;\lambda_t]$ as

$$\begin{aligned}\hat{S}[t;\lambda_t] &\equiv -\int d\Sigma_{t\mu}\left[\hat{T}^\mu{}_\nu(t,\boldsymbol{x})\beta^\nu(t,\boldsymbol{x}) + \hat{J}^{\mu\nu}(t,\boldsymbol{x})\mathcal{H}_\nu(t,\boldsymbol{x})\right] + \Psi[\lambda_t] \\ &= -\int d^3x\sqrt{-g}\left[\hat{T}^0{}_\nu(t,\boldsymbol{x})\beta^\nu(t,\boldsymbol{x}) + \hat{J}^{0\nu}(t,\boldsymbol{x})\mathcal{H}_\nu(t,\boldsymbol{x})\right] + \Psi[\lambda_t].\end{aligned} \quad (3.3)$$

Namely, we assume that the initial density operator take the form $\hat{\rho}_0 = \hat{\rho}_{\mathrm{LG}}[t_0;\lambda_{t_0}]$. In the following, we express an expectation value of an arbitrary operator $\hat{\mathcal{O}}$ over the local Gibbs distribution $\hat{\rho}_{\mathrm{LG}}[t;\lambda_t]$ as

$$\langle\hat{\mathcal{O}}\rangle_t^{\mathrm{LG}} \equiv \mathrm{Tr}\left(\hat{\rho}_{\mathrm{LG}}[t;\lambda_t]\hat{\mathcal{O}}\right). \quad (3.4)$$

The local Gibbs distribution, a generalization of the Gibbs distribution in a global equilibrium, fixes average values of the conserved charge densities with the help of conjugate Lagrange multipliers, which we collectively express as $\lambda_t \equiv \{\beta^\mu(t,\boldsymbol{x}),\mathcal{H}_\nu(t,\boldsymbol{x})\}$. Since the QED plasma supports the energy-momentum and magnetic flux as conserved charges, we introduced corresponding parameters $\beta^\mu$ and $\mathcal{H}_\nu$, which are identified as the fluid four-velocity $u^\mu$ and magnetic field $H_\mu$ multiplied by the local inverse temperature as $\beta^\nu = \beta u^\nu$ with $u^\nu u_\nu = -1$ and $\mathcal{H}_\mu = \beta H_\mu$. In this paper, we call the conjugate parameter $\mathcal{H}_\nu$ a "reduced magnetic field" after the presence of the inverse temperature prefactor. Since the zeroth component of the magnetic flux density identically vanishes (recall $\hat{J}^{00}=0$), the reduced magnetic field has only the spatial components; in other words, $\mathcal{H}_0=0$. Apart from a first-order phase transition, these Lagrange multipliers have one-to-one correspondences to the average conserved charge densities. Note that the first argument $t$ in Eqs. (3.2)-(3.3) corresponds to the operator argument, whereas the suffix $t$ for the parameter $\lambda$ specifies its configuration at time $t$. We also used the covariant description by introducing the space-like hypersurface $\Sigma_t$ with its normal vector $d\Sigma_{t\mu}$ proportional to a 3-dimensional volume element on the hypersurface (See, e.g., Refs. [63, 64] and explanation around Eq. (4.2) in detail).

To normalize the local Gibbs distribution as $\mathrm{Tr}\,\hat{\rho}_{\mathrm{LG}}[t;\lambda_t] = 1$, we introduced

$$\begin{aligned}\Psi[\lambda_t] &\equiv \log\mathrm{Tr}\exp\left(\int d\Sigma_{\bar{t}\mu}\left[\hat{T}^\mu{}_\nu(t,\boldsymbol{x})\beta^\nu(t,\boldsymbol{x}) + \hat{J}^{\mu\nu}(t,\boldsymbol{x})\mathcal{H}_\nu(t,\boldsymbol{x})\right]\right) \\ &= \log\mathrm{Tr}\exp\left(\int d^3x\sqrt{-g}\left[\hat{T}^0{}_\nu(t,\boldsymbol{x})\beta^\nu(t,\boldsymbol{x}) + \hat{J}^{0\nu}(t,\boldsymbol{x})\mathcal{H}_\nu(t,\boldsymbol{x})\right]\right).\end{aligned} \quad (3.5)$$

This functional, defined in the *local* thermal equilibrium state, serves as a generalization of the thermodynamic potential for the global equilibrium state and is called the Massieu-Planck functional. As in a global thermal equilibrium case, taking a variation with respect to the parameter $\lambda$, one can obtain the local Gibbs average of the conserved charge densities:

$$\langle\hat{T}^0{}_\mu(t,\boldsymbol{x})\rangle_t^{\mathrm{LG}} = \frac{1}{\sqrt{-g}}\frac{\delta\Psi[\lambda_t]}{\delta\beta^\mu(t,\boldsymbol{x})}, \quad \langle\hat{J}^{0\mu}(t,\boldsymbol{x})\rangle_t^{\mathrm{LG}} = \frac{1}{\sqrt{-g}}\frac{\delta\Psi[\lambda_t]}{\delta\mathcal{H}_\mu(t,\boldsymbol{x})}. \quad (3.6)$$

Thus, the Massieu-Planck functional keeps full information on the local Gibbs average of conserved charge densities. We shall again introduce a collective notation for the average charge densities as $c_t \equiv \{\langle \hat{T}^0{}_\mu(t,\boldsymbol{x})\rangle^{\text{LG}}_t, \langle \hat{J}^{0\mu}(t,\boldsymbol{x})\rangle^{\text{LG}}_t\}$[8]. Besides, one finds that the local Gibbs average of the entropy functional operator, $S[c_t] \equiv \text{Tr}\big(\hat{\rho}_{\text{LG}}[t;\lambda_t]\hat{S}[t;\lambda_t]\big)$, can be identified as the Legendre transformation from the parameter $\lambda_t$ to the average charge densities $c_t$:

$$S[c_t] = -\int \mathrm{d}^3 x \sqrt{-g}\left[\langle \hat{T}^0{}_\nu(t,\boldsymbol{x})\rangle^{\text{LG}}_t \beta^\nu(t,\boldsymbol{x}) + \langle \hat{J}^{0\nu}(t,\boldsymbol{x})\rangle^{\text{LG}}_t \mathcal{H}_\nu(t,\boldsymbol{x})\right] + \Psi[\lambda_t]. \quad (3.7)$$

Accordingly, we obtain a direct relation between the value of the parameter $\lambda_t$ and the average charge density—$T^0{}_\mu(t,\boldsymbol{x}) = \langle \hat{T}^0{}_\mu(t,\boldsymbol{x})\rangle^{\text{LG}}_t$ and $J^{0\mu}(t,\boldsymbol{x}) = \langle \hat{J}^{0\mu}(t,\boldsymbol{x})\rangle^{\text{LG}}_t$—as

$$\beta^\mu(t,\boldsymbol{x}) = -\frac{1}{\sqrt{-g}}\frac{\delta S[c_t]}{\delta T^0{}_\mu(t,\boldsymbol{x})}, \quad \mathcal{H}_\mu(t,\boldsymbol{x}) = -\frac{1}{\sqrt{-g}}\frac{\delta S[c_t]}{\delta J^{0\mu}(t,\boldsymbol{x})}, \quad (3.8)$$

whose global equilibrium limit matches with the definition of thermodynamic parameters following from the thermodynamic entropy. This is the reason why the functional operator $\hat{S}[t;\lambda_t]$ is referred to as the entropy operator.

## 3.2 Setup of the problem

Let us now set up the problem. We first define an expectation value of a Heisenberg operator $\hat{\mathcal{O}}(t)$ at an arbitrary time $t$:

$$\langle \hat{\mathcal{O}}(t)\rangle = \text{Tr}\left[\hat{\rho}_0 \hat{\mathcal{O}}(t)\right], \quad (3.9)$$

where we employed the Heisenberg picture, and thus, the density operator $\hat{\rho}_0$ does not evolve in time. The above assumption on the initial local Gibbs state will be explicit if one writes $\langle \hat{\mathcal{O}}(t)\rangle = \langle \hat{\mathcal{O}}(t)\rangle^{\text{LG}}_{t_0}$. Then, taking the expectation values of the operator identities (3.1), we obtain conservation laws for the averaged quantities at an arbitrary time $t$ ($\geq t_0$):

$$\nabla_\mu \langle \hat{T}^\mu{}_\nu(t,\boldsymbol{x})\rangle = \frac{1}{2}\langle \hat{J}^{\alpha\beta}(t,\boldsymbol{x})\rangle H_{\nu\alpha\beta}(t,\boldsymbol{x}), \quad \nabla_\mu \langle \hat{J}^{\mu\nu}(t,\boldsymbol{x})\rangle = 0. \quad (3.10)$$

The goal of this paper is to show that these equations finally result in a set of magnetohydrodynamic equations. To be more specific, we will express the spatial components of currents as functionals of the expectation values of conserved charge densities, so that the averaged conservation laws (3.1) form a closed set of equations. Those relations are called the constitutive relations:

$$\langle \hat{T}^\mu{}_\nu(t,\boldsymbol{x})\rangle = T^\mu{}_\nu\big[\langle \hat{T}^0{}_\nu\rangle, \langle \hat{J}^{0\nu}\rangle\big], \quad \langle \hat{J}^{\mu\nu}(t,\boldsymbol{x})\rangle = J^{\mu\nu}\big[\langle \hat{T}^0{}_\nu\rangle, \langle \hat{J}^{0\nu}\rangle\big], \quad (3.11)$$

where the right-hand side represents functional dependences of the average currents on the average charge densities. Thus, one of our main tasks is to derive the constitutive relations from the field-theoretical framework. We will explain that the derivation of constitutive relations is far from trivial even if we put a crucial assumption on the local thermal equilibrium for the initial state.

Before presenting challenging points of the problem, we shall check the numbers of the dynamical equations and degrees of freedom. For the averaged conservation laws (3.10) to form a closed set of hydrodynamic equations, the number of independent equations should match that of the conserved charges $\{\langle\hat{T}^0{}_\nu\rangle, \langle\hat{J}^{0\nu}\rangle\}$, namely seven (recall that $\hat{J}^{\mu\nu}$ is anti-symmetric so that $\hat{J}^{00} = 0$). One, however, finds that the temporal component of the second equation, $\nabla_\mu \langle\hat{J}^{\mu 0}\rangle = 0$, is a nondynamical constraint equation, implying an apparent mismatch between the numbers of dynamical equations and independent variables. However, it is crucial to notice that we have a trivial identity $\nabla_\nu(\nabla_\mu \langle\hat{J}^{\mu\nu}\rangle) \propto R_{\mu\nu}\langle\hat{J}^{\mu\nu}\rangle = 0$ (we used the fact that Ricci tensor $R_{\mu\nu}$ is symmetric: $R_{\mu\nu} = R_{\nu\mu}$). Therefore, both the numbers of independent equations and of conserved charges are six, and indeed match each other. In the following, we do not explicitly solve the constraint equation.

---

[8]We note $c_t$ does not include the Lapse function $N$ defined in Eq. (4.3), which is slightly different from the notation previously used in Refs. [63–65].

### 3.3 Time evolution as renormalized/optimized perturbation theory

We here explain challenging points of the problem and sketch our strategy to derive the constitutive relations. At the initial time $t_0$, the expectation values are simply specified by the initial local Gibbs distribution $\hat{\rho}_{\mathrm{LG}}[t_0; \lambda_{t_0}]$ and are provided as functionals of the thermodynamic parameters $\lambda_{t_0}$ at that time. Recalling the one-to-one correspondences between the values of parameters $\lambda$ and those of charge densities $T^0{}_\nu$ and $J^{0\nu}$, one can deduce the presence of constitutive relations (3.11) though its explicit form needs further investigation. At time $t\ (> t_0)$ of our interest in the future, the situation becomes more complicated. If we express the average currents $\langle\hat{T}^{\mu\nu}(t,\boldsymbol{x})\rangle$ and $\langle\hat{J}^{\mu\nu}(t,\boldsymbol{x})\rangle$ by applying the derivative expansion of the parameter at the initial time $t_0$, we soon encounter a meaningless divergent result due to a lack of the backreaction to thermodynamic variables (currents endlessly flow according to the initial parameter profile). Then, one would expect that the average currents $\langle\hat{T}^{\mu\nu}(t,\boldsymbol{x})\rangle$ and $\langle\hat{J}^{\mu\nu}(t,\boldsymbol{x})\rangle$ are expressed by a new parameter set $\lambda_t = \{\beta^\mu(t,\boldsymbol{x}), \mathcal{H}_\mu(t,\boldsymbol{x})\}$ at time $t$ instead of the initial one $\lambda_0$. However, the density operator in the Heisenberg picture does not evolve in time and, furthermore, we have not yet defined such a new parameter set.

To resolve the above difficulty, we turn our attention to what we practically do in solving hydrodynamic equations. Suppose that one knows the configuration of charge densities at time $t$ and, moreover, is equipped with constitutive relations in terms of the inverse temperature, fluid velocity, and magnetic field. Then, one can solve the hydrodynamic equations within an infinitesimal time step $\mathrm{d}t$ to obtain a new configuration of the average charge densities, $\langle\hat{T}^0{}_\nu(t+\mathrm{d}t,\boldsymbol{x})\rangle$ and $\langle\hat{J}^{0\nu}(t+\mathrm{d}t,\boldsymbol{x})\rangle$. To keep running hydrodynamic equations with the same algorithm that relies on the constitutive relations in terms of the parameter $\lambda$, we then need to feedback information on $\langle\hat{T}^0{}_\nu(t+\mathrm{d}t,\boldsymbol{x})\rangle$ and $\langle\hat{J}^{0\nu}(t+\mathrm{d}t,\boldsymbol{x})\rangle$ to update the corresponding parameters $\lambda_{t+\mathrm{d}t}$ (or define $\lambda_{t+\mathrm{d}t}$ from the average charge densities). One standard way is to use the local thermodynamic relation, which enable us to relates the charge densities to their conjugate parameters. From our perspective, this assumption can be interpreted as the matching condition for the full average charge density to a "hypothetical" local Gibbs average[9]:

$$\langle\hat{T}^0{}_\nu(t+\mathrm{d}t,\boldsymbol{x})\rangle = \langle\hat{T}^0{}_\nu(t+\mathrm{d}t,\boldsymbol{x})\rangle^{\mathrm{LG}}_{t+\mathrm{d}t}, \quad \langle\hat{J}^{0\nu}(t+\mathrm{d}t,\boldsymbol{x})\rangle = \langle\hat{J}^{0\nu}(t+\mathrm{d}t,\boldsymbol{x})\rangle^{\mathrm{LG}}_{t+\mathrm{d}t}.$$

---

[9]While there could be a room for improving the matching condition, we have not found superior one to the best of our knowledge.

The right-hand sides of these equations contain a new parameter set $\lambda_{t+\mathrm{d}t}$, which one can determine with the left-hand sides which have been obtained by solving the hydrodynamic equation. Note that the number of the matching condition coincides with that of parameters $\lambda(t)$. This matching condition completes our algorithm to solve the averaged conservation laws relying on the constitutive relations expressed by the local thermodynamic parameter $\lambda$. In short, we generally employ the matching condition

$$\langle \hat{T}^0{}_\nu(t,\boldsymbol{x})\rangle = \langle \hat{T}^0{}_\nu(t,\boldsymbol{x})\rangle^{\mathrm{LG}}_t, \quad \langle \hat{J}^{0\nu}(t,\boldsymbol{x})\rangle = \langle \hat{J}^{0\nu}(t,\boldsymbol{x})\rangle^{\mathrm{LG}}_t \quad \text{for} \quad \forall\, t \geq t_0. \tag{3.12}$$

These conditions enable us to define the local thermodynamic parameters $\lambda_t$ at a later time $t\ (>t_0)$.

Based on the new parameter set $\lambda_t$, we are now ready to explain how we express the average currents $\langle \hat{T}^{\mu\nu}(t,\boldsymbol{x})\rangle$ and $\langle \hat{J}^{\mu\nu}(t,\boldsymbol{x})\rangle$ in terms of $\lambda_t$. As we mentioned, the initial density operator does not evolve in time in the Heisenberg picture, and it is not clear how the new parameter set $\lambda_t$ manifests itself in expectation values of the conserved charge densities. However, a simple trick of the renormalized/optimized perturbation [71–73] resolves the problem: we formally rewrite the initial density operator as

$$\hat{\rho}_0 = \exp\big(-\hat{S}[t;\lambda_t] + \hat{\Sigma}[t,t_0;\lambda]\big) \quad \text{with} \quad \hat{\Sigma}[t,t_0;\lambda] \equiv \hat{S}[t;\lambda_t] - \hat{S}[t_0;\lambda_{t_0}]. \tag{3.13}$$

Note that the operator $\hat{\Sigma}[t,t_0;\lambda]$ gives the entropy difference between time $t$ and $t_0$, so that we will call it the entropy production operator. This arrangement is always allowed since the right-hand side is just identical to the original definition of $\hat{\rho}_{\mathrm{LG}}[t_0;\lambda_{t_0}]$. Keeping the noncommutative properties of the operators in mind, we factorize the exponential factor as

$$\hat{\rho}_0 = \exp\big(-\hat{S}[t;\lambda_t]\big)\hat{U}[t,t_0;\lambda] \quad \text{with} \quad \hat{U}[t,t_0;\lambda] \equiv \mathrm{T}_\tau \exp\left(\int_0^1 \mathrm{d}\tau\, \hat{\Sigma}_\tau[t,t_0;\lambda]\right), \tag{3.14}$$

where $\mathrm{T}_\tau$ denotes the time-ordering operator for $\tau$. We defined a shorthand notation $\hat{\mathcal{O}}_\tau \equiv \mathrm{e}^{\tau \hat{S}[t;\lambda]} \hat{\mathcal{O}} \mathrm{e}^{-\tau \hat{S}[t;\lambda]}$ for an arbitrary operator $\hat{\mathcal{O}}$. Thanks to this rearranged factorization, we obtain an identity, which expresses the expectation value of an arbitrary operator $\hat{\mathcal{O}}(t)$ at time $t$ by the local Gibbs average at the same time:

$$\langle \hat{\mathcal{O}}(t)\rangle = \mathrm{Tr}\left(\mathrm{e}^{-\hat{S}[t;\lambda_t]}\hat{U}[t,t_0;\lambda]\hat{\mathcal{O}}(t)\right) = \langle \hat{U}[t,t_0;\lambda]\hat{\mathcal{O}}(t)\rangle^{\mathrm{LG}}_t. \tag{3.15}$$

This observation motivates us to evaluate $\langle \hat{\mathcal{O}}(t)\rangle$ by expanding the exponential operator $\hat{U}[t,t_0;\lambda]$, which leads to an expansion on top of the "hypothetical" local equilibrium state at each time slice. Namely, one could expand the "evolution operator" $\hat{U}[t,t_0;\lambda]$ with respect to $\hat{\Sigma}[t,t_0;\lambda]$, in which the leading term reduces to the thermal expectation value $\langle \hat{\mathcal{O}}(t)\rangle^{\mathrm{LG}}_t$ in line with our expectation. This prescription gives the renormalized/optimized perturbation theory for the hydrodynamic time evolution.

In order for this renormalized/optimized peturbation to work, the entropy production operator $\hat{\Sigma}[t,t_0;\lambda]$ needs to be a small quantity so that it works as a controlled expansion parameter. One can confirm this is indeed the case in the sense of the derivative expansion.

By rewriting the entropy production operator as

$$\hat{\Sigma}[t, t_0; \lambda] = \int_{t_0}^{t} dt' \partial_{t'} \hat{S}[t'; \lambda_{t'}]$$
$$= -\int_{t_0}^{t} dt' \int d^3x \sqrt{-g} \left[ \frac{1}{2} \delta \hat{T}^{\mu\nu} (\nabla_\mu \beta_\nu + \nabla_\nu \beta_\mu) + \frac{1}{2} \delta \hat{J}^{\mu\nu} \left( \beta^\rho H_{\rho\mu\nu} + (\nabla_\mu \mathcal{H}_\nu - \nabla_\nu \mathcal{H}_\mu) \right) \right],$$
(3.16)

one finds that the right-hand side consists only of the parameter (and background gauge field) derivatives. Here, we introduced $\delta \hat{\mathcal{O}}(t) \equiv \hat{\mathcal{O}}(t) - \langle \hat{\mathcal{O}}(t) \rangle_t^{\mathrm{LG}}$ for an arbitrary Heisenberg operator $\hat{\mathcal{O}}(t)$. We also used the Stokes' theorem $\partial_t \int d\Sigma_{t\mu} f^\mu = \int d^3x \sqrt{-g} \nabla_\mu f^\mu$ and the covariant conservation laws (3.1), together with the following expression for the time derivative of the Massieu-Planck functional $\Psi[\lambda_t]$:

$$\partial_t \Psi[\lambda_t] = \int d^3x \sqrt{-g} \left[ \frac{1}{2} \langle \hat{T}^{\mu\nu} \rangle_t^{\mathrm{LG}} (\nabla_\mu \beta_\nu + \nabla_\nu \beta_\mu) + \frac{1}{2} \langle \hat{J}^{\mu\nu} \rangle_t^{\mathrm{LG}} \left( \beta^\rho H_{\rho\mu\nu} + (\nabla_\mu \mathcal{H}_\nu - \nabla_\nu \mathcal{H}_\mu) \right) \right].$$
(3.17)

The subtraction of $\partial_{t'} \Psi[\lambda_{t'}]$ leads to displacements $\delta \hat{T}^{\mu\nu}$ and $\delta \hat{J}^{\mu\nu}$ in Eq. (3.16). The derivative quantities on the right-hand side of Eq. (3.16) are expected to take small values when hydrodynamics works in describing the low-energy behavior (i.e., low-frequency and long-wavelength limits) of conserved charge densities. In such a situation, one may systematically organize a *derivative expansion* of the constitutive equations via the expansion of $\hat{U}[t, t_0; \lambda]$ with respect to $\hat{\Sigma}[t, t_0; \lambda]$.

Our basic strategy has been implemented with the above rearrangements. Based on those crucial observations with introduction of $\hat{U}[t, t_0; \lambda]$, we now decompose the problem into two separate parts: the local equilibrium average and deviation from it. This decomposition is accomplished by arranging the average currents as

$$\langle \hat{T}^\mu_\nu(t, \boldsymbol{x}) \rangle = \langle \hat{T}^\mu_\nu(t, \boldsymbol{x}) \rangle_t^{\mathrm{LG}} + \langle \delta \hat{T}^\mu_\nu(t, \boldsymbol{x}) \rangle, \quad (3.18\mathrm{a})$$
$$\langle \hat{J}^{\mu\nu}(t, \boldsymbol{x}) \rangle = \langle \hat{J}^{\mu\nu}(t, \boldsymbol{x}) \rangle_t^{\mathrm{LG}} + \langle \delta \hat{J}^{\mu\nu}(t, \boldsymbol{x}) \rangle, \quad (3.18\mathrm{b})$$

where the displacements are given by

$$\langle \delta \hat{T}^\mu_\nu(t, \boldsymbol{x}) \rangle = \langle (\hat{U}[t, t_0; \lambda] - 1) \hat{T}^\mu_\nu(t, \boldsymbol{x}) \rangle_t^{\mathrm{LG}}, \quad (3.19\mathrm{a})$$
$$\langle \delta \hat{J}^{\mu\nu}(t, \boldsymbol{x}) \rangle = \langle (\hat{U}[t, t_0; \lambda] - 1) \hat{J}^{\mu\nu}(t, \boldsymbol{x}) \rangle_t^{\mathrm{LG}}. \quad (3.19\mathrm{b})$$

Those terms contain at least one $\hat{\Sigma}[t, t_0; \lambda]$, and thus, lead to derivative corrections to the constitutive relations. In the subsequent sections, we will separately investigate the first and second terms of Eq. (3.18) in detail to derive the constitutive relations for relativistic magnetohydrodynamics.

Before closing this section, we provide a comment on the matching condition used to update the parameter set $\lambda$ in the time increment. Using the above notations, one can rewrite the matching conditions as $\langle \delta \hat{T}^0_\nu(t, \boldsymbol{x}) \rangle = 0$ and $\langle \delta \hat{J}^{0\nu}(t, \boldsymbol{x}) \rangle = 0$, indicating the absence of the derivative corrections to conserved charges. However, this does not mean that the system is in a local equilibrium state at every time slice $t$ since the density operator is always given by the initial local Gibbs distribution. Putting it differently, the matching condition only means that we have matched the conserved charge densities at time $t$ to those in the hypothetical local equilibrium forms by adjusting the parameter set $\lambda_t$ (note that the full averages of other operators do not agree with the newly introduced local Gibbs average of those operators). Thus, it is legitimate to say that introduction of a new parameter set $\lambda_t$ with the matching condition (3.12) should be understood as one particular approximation scheme such as the one employed in the renormalized/optimized perturbation theory [71–73]. From this viewpoint, one can identify the matching condition (3.12) as a counterpart of the fastest apparent convergence condition used in the optimized perturbation theory [71]. It is also worth mentioning that our matching condition (3.12) with the decomposition $\beta^\mu = \beta u^\mu$ ($u^\mu u_\mu = -1$) leads to one natural definition of a "fluid-velocity" $u^\mu$, which is known as that defined in the so-called beta frame [62]. This fluid velocity is, in general, different from that of the Landau-Lifshitz frame (or, of courase, that of the Eckart frame). This "frame choice" is usually implicitly assumed in the phenomenological derivation of relativistic hydrodynamics, but is not at all a unique choice (see, e.g., Refs. [74, 75]).

## 4 Constitutive relations: Exact results

In this section, we derive an exact formula for the constitutive equations (3.11) for both the nondissipative and dissipative components. Before starting a discussion, we briefly summarize our notations for the coordinate system (see Refs. [63, 64, 76] for more systematic explanations). While we have introduced a covariant notation to describe constant time hypersurface $\Sigma_t$ in the previous section, other parts of the coordinate system have been left implicit. In order to make our formulation to be manifestly covariant under diffeomorphism, we put the QED plasma in a curved spacetime equipped with the background vierbein (metric) as before. However, we emphasize that even in a flat spacetime there is a benefit to maintain the general covariance by using the curvlinear coordinate system—as was the case in the covariant formulation (super-many-time theory) of QED [1–8]—so that we can choose a suitable coordinate system to describe the *local* equilibrium state with a non-uniform flow velocity.

Let $\bar{x}^{\bar{\mu}}$ be our choice of the coordinate system and assume existence of an inverse mapping $x^\mu(\bar{x})$ to another coordinate system $x^\mu = (t, \boldsymbol{x})$ in general. Using a scalar time function $\bar{t}(x)$, we introduce a foliation speficying equal-time (or constant time) hypersurfaces $\Sigma_{\bar{t}}$, on each of which the time function takes a constant value: $\bar{t}(x) = $ const. Expressing its component as $\bar{x}^{\bar{\mu}}(x) \equiv \big(\bar{t}(x), \bar{\boldsymbol{x}}(x)\big)$, one finds $\bar{\boldsymbol{x}}$ defines the coordinate on the spatial hypersurface $\Sigma_{\bar{t}}$ at each time slice parametrized by a value of the time coordinate $\bar{t}$. Then, we define two timelike vectors which specify the time direction of the local coordinate system and the normal direction to the hypersurface $\Sigma_{\bar{t}}$. Explicitly, they are given by

$$t^\mu \equiv \partial_{\bar{t}} x^\mu(\bar{x}) \quad \text{and} \quad n_\mu \equiv -N \partial_\mu \bar{t}(x). \qquad (4.1)$$

The time vector $t^\mu$ and the normal vector $n^\mu$ are proportional to each other if we use the Cartesian coordinate system. The normal vector has a unit norm $n^\mu n_\mu = -1$, and the normalization function $N(x)$ is called the lapse function. On the other hand, the time vector is not necessarily normalized, and its norm represents how we measure the scale along local time directions. We also require that the normal vector satisfy a condition $n \wedge dn = 0$ so that the hypersurface covers all the space without intersecting another hypersurface thanks to the Frobenius theorem. By using the normal vector, we can write the infinitesimal surface element vector as

$$d\Sigma_{\bar{t}\mu} \equiv -d\Sigma_{\bar{t}} n_\mu = -d^3 \bar{x} \sqrt{\gamma} n_\mu. \tag{4.2}$$

On the rightmost side, we introduced an (invariant) spatial volume element on the hypersurface by using the determinant of the induced metric: $\gamma = \det \gamma_{\mu\nu}$ with $\gamma_{\mu\nu} \equiv g_{\mu\nu} + n_\mu n_\nu$.

Decomposing the time vector with the normal vector $n^\mu$ and its orthogonal component, we have

$$t^\mu = N n^\mu + N^\mu \quad \text{with} \quad n_\mu N^\mu = 0, \tag{4.3}$$

where $N^\mu$ is called a shift vector. Using this $(3+1)$-decomposition, we can express the (inverse) metric $g_{\bar{\mu}\bar{\nu}}$ ($g^{\bar{\mu}\bar{\nu}}$) in the coordinate system $\bar{x}^\mu$ as

$$g_{\bar{\mu}\bar{\nu}} = \begin{pmatrix} -N^2 + N_{\bar{i}} N^{\bar{i}} & N_{\bar{i}} \\ N_{\bar{i}} & \gamma_{\bar{i}\bar{j}} \end{pmatrix}, \quad g^{\bar{\mu}\bar{\nu}} = \begin{pmatrix} -N^{-2} & N^{-2} N^{\bar{j}} \\ N^{-2} N^{\bar{i}} & \gamma^{\bar{i}\bar{j}} - N^{-2} N^{\bar{i}} N^{\bar{j}} \end{pmatrix}, \tag{4.4}$$

where we introduced $N_{\bar{i}} = \gamma_{\bar{i}\bar{j}} N^{\bar{j}}$ and the inverse $\gamma^{\bar{i}\bar{j}}$ of the induced metric satisfying $\gamma_{\bar{i}\bar{k}} \gamma^{\bar{k}\bar{j}} = \delta_{\bar{i}}^{\bar{j}}$. Note that the determinant of spacetime metric $g_{\mu\nu}$ is given by product of the Lapse $N$ and spatial determinant $\gamma$ as $\sqrt{-g} = N\sqrt{\gamma}$. Since we will basically use only the coordinate system of $\bar{x}^\mu$ in the rest of this paper, we will omit all the overbars to express our coordinate system $\bar{x}^\mu$, which is thus simply denoted as $x^\mu$.

## 4.1 Nondissipative part from the Massieu-Planck functional

In this subsection, we examine the nondissipative (or local equilibrium) part given by the first term in Eq. (3.18). First, we show that the Massieu-Planck functional (3.5) works as a generating functional of the conserved currents—not only for the charge density as given in Eq. (3.6). This means that all the transport phenomena taking place in the local equilibrium state could be derived from one functional $\Psi[\lambda_t]$. Then, we provide a path-integral representation of the Massieu-Planck functional, and show that the local equilibrium configuration can be generally interpreted as an emergent curved spacetime geometry/two-form gauge connection in the imaginary-time (or Matsubara) formalism [77, 78]. Notion of thermally induced curved spacetime/gauge connection, in particular its symmetry properties, allows us to give a systematic construction of the Massieu-Planck functional. Since this formalism can be constructed in parallel to previous works [63, 64, 66], we will highlight differences from the previous works that are induced by the one-form symmetry newly introduced in this work.

### 4.1.1 Massieu-Planck functional as generating functional

**Derivation of variational formulas.** Let us show that the Massieu-Planck functional serves as a generating functional for the local Gibbs average so that it contains all the information on the local equilibrium transport. For notational brevity, we shall introduce a functional operator $\hat{K}[t;\lambda]$ (an operator part of the entropy $\hat{S}[t;\lambda]$) as

$$\hat{K}[t;\lambda,j] \equiv -\int d\Sigma_{t\mu} \left[ \hat{T}^\mu_{\ \nu}(t,\boldsymbol{x})\beta^\nu(t,\boldsymbol{x}) + \hat{J}^{\mu\nu}(t,\boldsymbol{x})\mathcal{H}_\nu(t,\boldsymbol{x}) \right], \qquad (4.5)$$

which leads to a compact expression of the Massieu-Planck functional as $\Psi[t;\lambda,j] = \log \mathrm{Tr}\, e^{-\hat{K}[t;\lambda,j]}$. Since the background field $j = \{e_\mu^{\ a}, b_{\mu\nu}\}$ plays a crucial role in the following discussion, we explicitly show it in the functional argument.

On the basis of this definition, let us consider a variation of the Massieu-Planck functional under an infinitesimal transformation acting on both the local Gibbs parameter $\lambda$ and background field $j$. To be explicit, we consider a combination of the infinitesimal general coorinate transformation and 1-form gauge transformation acting on them as

$$\epsilon\delta_\lambda \lambda = \pounds_\xi \lambda, \quad \epsilon\delta_\lambda e_\mu^{\ a} = \pounds_\xi e_\mu^{\ a}, \quad \epsilon\delta_\lambda b_{\mu\nu} = \pounds_\xi b_{\mu\nu} + \partial_\mu \theta_\nu - \partial_\nu \theta_\mu, \quad \epsilon\delta_\lambda t = \pounds_\xi t, \qquad (4.6)$$

with a specific choice of the transformation parameter $\xi^\mu = \epsilon\beta^\mu$ and $\theta_\mu = \epsilon(\mathcal{H}_\mu - \beta^\nu b_{\nu\mu})$ ($\epsilon$ denotes an infinitesimal constant). Note that the background two-form gauge field has a contribution from the one-form gauge transformation, and that the time function $t(x)$ specifying the equal-time hypersurface also transforms under the general coordinate transformation. Direct computation leads to the following explicit forms:

$$\begin{cases} \delta_\lambda \beta^\mu = 0, \\ \delta_\lambda \mathcal{H}_\mu = \beta^\rho \delta_\lambda b_{\rho\mu} + \partial_\mu(\beta^\rho \mathcal{H}_\rho), \\ \delta_\lambda e_\mu^{\ a} = -\beta^\nu \omega_\nu^{\ a}{}_b e_\mu^{\ b} + e_\nu^{\ a}\nabla_\mu\beta^\nu, \\ \delta_\lambda b_{\mu\nu} = \beta^\rho H_{\rho\mu\nu} + \nabla_\mu \mathcal{H}_\nu - \nabla_\nu \mathcal{H}_\mu, \\ \delta_\lambda t = N^{-1}\beta', \end{cases} \qquad (4.7)$$

where we used the vierbein postulate $\nabla_\mu e_\nu^{\ a} + \omega_\mu^{\ a}{}_b e_\nu^{\ b} = 0$ for the third equation, and defined $\beta' \equiv -\beta^\mu n_\mu$ for the last equation.

From the definition of the Massieu-Planck functional ($\Psi[t;\lambda,j] = \log\mathrm{Tr}\, e^{-\hat{K}[t;\lambda,j]}$), one immediately finds a simple variational relation $\delta_\lambda \Psi[t;\lambda,j] = -\langle \delta_\lambda^{\mathrm{para}} \hat{K}[t;\lambda,j]\rangle_t^{\mathrm{LG}}$, where $\delta_\lambda^{\mathrm{para}}$ means that the transformation only acts on the paramters in $\hat{K}[t;\lambda,j]$. The vital point here is that the functional operator $\hat{K}[t;\lambda,j]$ remains invariant under the simultaneous transformation acting both on the parameter and operator; namely, $\delta_\lambda^{\mathrm{tot}} \hat{K} \equiv \delta_\lambda^{\mathrm{para}}\hat{K} + \delta_\lambda^{\mathrm{op}}\hat{K} = 0$. Since the operator variation is generated by taking a self commutation relation, it identically vanishes: $\delta_\lambda^{\mathrm{op}}\hat{K} = [\mathrm{i}\hat{K},\hat{K}] = 0$. Therefore, the above variational relation leads to the following identity for the Massieu-Planck functional:

$$\delta_\lambda \Psi[t;\lambda,j] = -\langle \delta_\lambda^{\mathrm{para}}\hat{K}[t;\lambda,j]\rangle_t^{\mathrm{LG}} = 0, \qquad (4.8)$$

Let us then specify an explicit form of $\delta_\lambda \Psi[t;\lambda,j]$ and derive the variational formula. Recalling that $\Psi[t;\lambda,j]$ depends on the time function $t(x)$, which also transforms as a scalar, we evaluate the variation of the Massieu-Planck functional with respect to the time function as follows:

$$\frac{\delta\Psi}{\delta t(x)} = -\left\langle \frac{\delta\hat{K}}{\delta t(x)} \right\rangle_t^{\mathrm{LG}} = -\sqrt{-g}\left[ \langle \hat{T}_a^{\ \mu}(x)\rangle_t^{\mathrm{LG}} \delta_\lambda e_\mu^{\ a}(x) + \frac{1}{2}\langle \hat{J}^{\mu\nu}(x)\rangle_t^{\mathrm{LG}} \delta_\lambda b_{\mu\nu}(x)\right], \qquad (4.9)$$

– 15 –

which agrees with the integrand of Eq. (3.17). To show the second equality, we formally expressed $\hat{K}[t;\lambda,j]$ as a spacetime integral with the Heaviside step function $\theta(x)$ as [64, 66]

$$\hat{K}[t;\lambda,j] = \int d^4x \sqrt{-g}\,\theta\big(t-\bar{t}(x)\big)\left[\hat{T}^\mu_\nu \nabla_\mu \beta^\nu + \frac{1}{2}\hat{J}^{\mu\nu}(\beta^\rho H_{\rho\mu\nu} + \nabla_\mu \mathcal{H}_\nu - \nabla_\nu \mathcal{H}_\mu)\right], \quad (4.10)$$

where we used the conservation laws at the operator level (3.1). Then, using the transformation rule (4.7) together with the operator version of the constraint (2.11), we obtained the rightmost side of Eq. (4.9) (see Refs. [64, 66] for more details). Taking into account that contribution, we obtain $\delta_\lambda \Psi$ as follows:

$$\delta_\lambda \Psi = \int d^3x \left[\frac{\delta\Psi}{\delta t(x)}\delta_\lambda t(x) + \frac{\delta\Psi}{\delta\lambda^a(x)}\delta_\lambda \lambda^a(x) + \frac{\delta\Psi}{\delta e_\mu^{\ a}(x)}\delta_\lambda e_\mu^{\ a}(x) + \frac{\delta\Psi}{\delta b_{\mu\nu}(x)}\delta_\lambda b_{\mu\nu}(x)\right]$$

$$= \int d^3x \bigg[\sqrt{\gamma}\beta'\left(-\langle \hat{T}^\mu_a\rangle_t^{\mathrm{LG}} \delta_\lambda e_\mu^{\ a} - \frac{1}{2}\langle \hat{J}^{\mu\nu}\rangle_t^{\mathrm{LG}} \delta_\lambda b_{\mu\nu}\right)$$

$$+ \frac{\delta\Psi}{\delta \mathcal{H}_\nu}[\beta^\mu \delta_\lambda b_{\mu\nu} + \partial_\nu(\beta^\mu \mathcal{H}_\mu)] + \frac{\delta\Psi}{\delta e_\mu^{\ a}}\delta_\lambda e_\mu^{\ a} + \frac{\delta\Psi}{\delta b_{\mu\nu}}\delta_\lambda b_{\mu\nu}\bigg]$$

$$= -\int d^3x \left[\left(\sqrt{\gamma}\beta'\langle \hat{T}^\mu_a\rangle_t^{\mathrm{LG}} - \frac{\delta\Psi}{\delta e_\mu^{\ a}}\right)\delta_\lambda e_\mu^{\ a} + \left(\frac{1}{2}\sqrt{\gamma}\beta'\langle \hat{J}^{\mu\nu}\rangle_t^{\mathrm{LG}} - \frac{\delta\Psi}{\delta b_{\mu\nu}} - \beta^\mu \frac{\delta\Psi}{\delta \mathcal{H}_\nu}\right)\delta_\lambda b_{\mu\nu}\right]. \quad (4.11)$$

To obtain the third line, recalling Eq. (3.6), we performed an integration by parts as

$$\int d^3x \frac{\delta\Psi}{\delta \mathcal{H}_\nu}\partial_\nu(\beta^\mu \mathcal{H}_\mu) = \int d^3x \sqrt{-g}\langle \hat{J}^{0\nu}\rangle_t^{\mathrm{LG}}\partial_\nu(\beta^\mu \mathcal{H}_\mu) = -\int d^3x \sqrt{-g}\beta^\mu \mathcal{H}_\mu \nabla_i \langle \hat{J}^{0i}\rangle_t^{\mathrm{LG}} = 0,$$

where we neglected the surface term and used the zeroth-component of the conservation law for $\hat{J}^{0\mu}$. As a consequence, the identity $\delta_\lambda \Psi = 0$ results in

$$\int d^3x \left[\left(\sqrt{\gamma}\beta'\langle \hat{T}^\mu_a\rangle_t^{\mathrm{LG}} - \frac{\delta\Psi}{\delta e_\mu^{\ a}}\right)\delta_\lambda e_\mu^{\ a} + \left(\frac{1}{2}\sqrt{\gamma}\beta'\langle \hat{J}^{\mu\nu}\rangle_t^{\mathrm{LG}} - \frac{\delta\Psi}{\delta b_{\mu\nu}} - \beta^\mu \frac{\delta\Psi}{\delta \mathcal{H}_\nu}\right)\delta_\lambda b_{\mu\nu}\right] = 0. \quad (4.12)$$

Here, the variations $\delta_\lambda e_\mu^{\ a}$ and $\delta_\lambda b_{\mu\nu}$ are still left arbitrary since any parameter configuration $\lambda$ should satisfy $\delta_\lambda \Psi = 0$ irrespective of the hydrodynamic evolution of $\beta^\mu$ and $\mathcal{H}_\mu$ in the transformation parameters. Therefore, we eventually find the following variational formulas:

$$\langle \hat{T}^\mu_a(t,\boldsymbol{x})\rangle_t^{\mathrm{LG}} = \frac{1}{\beta'\sqrt{\gamma}}\frac{\delta\Psi[\lambda_t]}{\delta e_\mu^{\ a}(t,\boldsymbol{x})}, \quad (4.13a)$$

$$\langle \hat{J}^{\mu\nu}(t,\boldsymbol{x})\rangle_t^{\mathrm{LG}} = \frac{2}{\beta'\sqrt{\gamma}}\left(\beta^{[\mu}\delta_\rho^{\nu]}\frac{\delta\Psi[\lambda_t]}{\delta \mathcal{H}_\rho(t,\boldsymbol{x})} + \frac{\delta\Psi[\lambda_t]}{\delta b_{\mu\nu}(t,\boldsymbol{x})}\right), \quad (4.13b)$$

where we defined a shorthand notation for anti-symmetrization as $A^{[\mu}B^{\nu]} = (A^\mu B^\nu - B^\nu A^\mu)/2$. It is worth emphasizing that this gives an exact formula without relying on the derivative expansion. Using those exact formulas, one can extract all the transport phenomena taking place in the local thermal equilibrium state from the single functional $\Psi[\lambda_t]$. The derivative expansion of each local equilibrium constitutive relation is reduced from that of the Massieu-Planck functional on an order-by-order basis. In short, construction of the constitutive relation for the nondissipative parts amounts to computing the Massieu-Planck functional $\Psi[\lambda_t]$.

**Hydrostatic gauge.** While we have derived the general variational formulas without specifying the coordinate system, there is a useful gauge (or choice of the coordinate system), which we call *hydrodstatic gauge* [64]. To get motivated to introduce the hydrostatic gauge, we turn our attention to the fact that each fluid element has its *local* rest frame. On the other hand, we have acquired the freedom to choose the local time-direction of our coordinate system thanks to the covariant formulation. Therefore, we can make a suitable choice of the local coordinate system, in that all the fluid elements in the system look as if static.

Here, we introduce the hydrostatic gauge and simplify the variational formulas (4.13). The hydrostatic gauge is defined by the following gauge conditions:

$$t^\mu(x)\big|_{\text{hs}} = \beta^\mu(x)/\beta_0, \quad b_{0\mu}(x)\big|_{\text{hs}} = \mathcal{H}_\mu(x)/\beta_0, \quad (4.14)$$

where we introduced an arbitrary reference inverse temperature $\beta_0$ which has the same mass dimension as that of the inverse temperature. Here, the subscripts stand for the quantities given in the hydrostatic gauge. The first condition for the time vector $t^\mu$ indicates that the local time direction is taken along the local fluid flow and the scale of the time axis is specified by the local inverse temperature [63]. Besides, we put the second condition to extend the hydrostatic gauge for magnetohydrodynamics, which means that we interpret the reduced magnetic field $\mathcal{H}_\mu$ as a temporal component of the background two-form gauge field[10].

Thanks to the hydrostatic condition, we obtain $\mathcal{H}_\mu(x) - \beta^\nu b_{\nu\mu}\big|_{\text{hs}} = 0$ so that one-form gauge transformation in Eq. (4.7) becomes trivial as $\theta_\nu\big|_{\text{hs}} = \epsilon(\mathcal{H}_\mu - \beta^\nu b_{\nu\mu})\big|_{\text{hs}} = 0$. Thus, we can simplify a little bit complicated transformation $\delta_\lambda$ in Eq. (4.7) to just the lie derivative along the time direction $\pounds_\beta = \beta_0 \pounds_t\big|_{\text{hs}}$. Furthemore, the variation with respect to $\mathcal{H}_\mu$ is absorbed into that with respect to $b_{\mu\nu}$ so that we only need to take account of the variation with respect to background fields. Therefore, the identity $\delta_\lambda \Psi = 0$ in the hydrostatic gauge takes the following simple form:

$$\int \mathrm{d}^3 x \left[ \left( \beta_0 \sqrt{-g} \langle \hat{T}^\mu_a \rangle^{\text{LG}}_t - \frac{\delta \Psi}{\delta e_\mu{}^a} \right) \pounds_\beta e_\mu{}^a + \left( \frac{1}{2} \beta_0 \sqrt{-g} \langle \hat{J}^{\mu\nu} \rangle^{\text{LG}}_t - \frac{\delta \Psi}{\delta b_{\mu\nu}} \right) \pounds_\beta b_{\mu\nu} \right] = 0, \quad (4.15)$$

where we used $\beta' = \beta_0 N\big|_{\text{hs}}$. Therefore, we obtain simpler variational formulas in the hydrostatic gauge

$$\langle \hat{T}^\mu_a(t,\boldsymbol{x}) \rangle^{\text{LG}}_t = \frac{1}{\beta_0 \sqrt{-g}} \frac{\delta \Psi[\lambda_t]}{\delta e_\mu{}^a(t,\boldsymbol{x})}\bigg|_{\text{hs}}, \quad \langle \hat{J}^{\mu\nu}(t,\boldsymbol{x}) \rangle^{\text{LG}}_t = \frac{2}{\beta_0 \sqrt{-g}} \frac{\delta \Psi[\lambda_t]}{\delta b_{\mu\nu}(t,\boldsymbol{x})}\bigg|_{\text{hs}}. \quad (4.16)$$

This expression means that the Massieu-Planck functional in the hydrostatic gauge serves as an analogue of the action functional, which implies a possibility of the action principle description of hydordynamics (See, e.g., Ref. [44] for recent discussions).

---

[10]This condition is completely in parallel with that of (zero-form) U(1) symmetry, where a local chemical potential $\mu$ is regarded as the temporal component of the background gauge field $A_0 = \beta(x)\mu(x)/\beta_0$ (See Ref. [64]).

### 4.1.2 Path-integral formalism for the Massieu-Planck functional

According to the variational formulas shown in the previous section, we only need to compute the Massieu-Planck functional to evaluate the local equilibrium averages of current operators. In the derivation of the variational formulas, we only used the consequences resulting from symmetries, specifically the spacetime and magnetic one-form symmetries, which are independent of microscopic details of the system.

We now specify the QED action (2.3) as the underlying microscopic theory[11], and derive the path-integral representation of the Massiue-Planck functional (3.5). We take the the operator (Hamiltonian) formalism on the basis of the $(3+1)$ decomposition [76] as the starting point. The relevant dynamical degrees of freedom are the sum of the Dirac field and U(1) gauge field separately considered in Ref. [64], so that we can apply the path-integral formulas shown therein. Putting an emphasis on the specific points arising in the presence of the magnetic one-form symmetry, we here sketch our derivation (see Ref. [64] for the detailed calculation).

**Derivation of path-integral formula.** We shall elaborate the QED action (2.3) equipped with the background field. Since the fermion sector is almost same as the one in Ref. [64], we here concentrate on the gauge field sector, which contains the two-form background field. The $(3+1)$ decomposition of the photon sector leads to the following Lagrangian:

$$\sqrt{-g}\mathcal{L}_{\text{QED}}^{\text{gauge}} \equiv -\frac{\sqrt{-g}}{4}g^{\mu\nu}g^{\alpha\beta}F_{\mu\alpha}F_{\nu\beta} + \frac{\sqrt{-g}}{2}b_{\mu\nu}\tilde{F}^{\mu\nu}$$
$$= \frac{\sqrt{\gamma}}{2N}(F_{0i} - N^k F_{ki})\gamma^{ij}(F_{0j} - N^l F_{lj}) - \frac{N\sqrt{\gamma}}{4}\gamma^{ij}\gamma^{kl}F_{ik}F_{jl} + \frac{1}{2}\epsilon^{0ijk}(b_{0i}F_{jk} + b_{jk}F_{0i}),$$
(4.17)

from which one can define the canonical momentum of the gauge field $A_i$ as

$$\Pi^i \equiv \frac{\partial(\sqrt{-g}\mathcal{L}_{\text{QED}})}{\partial(\partial_0 A_i)} = -\sqrt{-g}F^{0i} + \epsilon^{0ijk}b_{jk} = \frac{\sqrt{\gamma}}{N}\gamma^{ij}(F_{0j} - N^k F_{kj}) + \frac{1}{2}\epsilon^{0ijk}b_{jk}. \quad (4.18)$$

Thus, the presence of the background two-form field generates a c-number shift for the conjugate momentum, which does not affect the canonical commutation relation. It is useful to define a shifted momentum

$$\acute{\Pi}^i \equiv \Pi^i - \frac{1}{2}\epsilon^{0ijk}b_{jk}, \quad (4.19)$$

which satisfies the same commutation relation as $\Pi^i$ does.

On the other hand, we need to pay attention to the Gauss's law since the Maxwell equation contains the contribution from the source term:

$$0 = \frac{\delta \mathcal{S}_{\text{QED}}}{\delta A_\mu} = \nabla_\mu F^{\mu\nu} + j^\nu + \frac{1}{2}\nabla_\mu(\varepsilon^{\nu\mu\rho\sigma}b_{\rho\sigma}) \quad \text{with} \quad j^\mu(x) \equiv \frac{1}{\sqrt{-g}}\frac{\delta \mathcal{S}_{\text{QED}}^{\text{mat}}}{\delta A_\mu(x)} = \mathrm{i}q\bar{\psi}\gamma^\mu\psi,$$
(4.20)

---

[11]While we focus on QED in $(3+1)$ dimension, we expect that the emergence of the same thermal background and the geometrical interpretation of the local thermal equilibrium state holds in any system endowed with one-form symmetry.

where we introduced an electric current $j^\mu$ by a variation of the fermionic part of the QED action $\mathcal{S}_{\text{QED}}^{\text{mat}}$ with respect to the gauge field. Expressing the contribution from the background field as $j^\mu_{\text{bkd}}(x) \equiv \frac{1}{2}\nabla_\nu(\varepsilon^{\mu\nu\rho\sigma}b_{\rho\sigma})$, one can regard the insertion of the background two-form field as putting the background electric current (see also Ref. [35] for more discussions). The Gauss's law from the temporal component of the Maxwell equation reads

$$\nabla_i \Pi^i = -\sqrt{-g}j^0 \;\Leftrightarrow\; \nabla_i \acute{\Pi}^i = -\sqrt{-g}j^0 - \frac{1}{2}\nabla_i(\epsilon^{0ijk}b_{jk}), \qquad (4.21)$$

which contains a c-number shift if we use $\acute{\Pi}^i$. Note that the temporal component of the canonical momentum vanishes as usual, $\Pi^0 = 0$, due to the gauge redundancy, so that there is no contribution from the background field.

As a next step, let us write down the conserved charge densities based on their definitions (2.8). Recalling that the coupling term induced by the two-form external field is independent of the vierbein [see the second line of Eq. (4.17)], we find that the background two-form field does not contribute to the energy-momentum tensor [recall also the energy-momentum tensor (2.8) is defined via the variation of the action with respect to the vierbein]. Therefore, the energy-momentum tensor has exactly the same form as the conventional case. In terms of $\acute{\Pi}^i$ defined in Eq. (4.19), the explicit forms of the photon contribution to the conserved charge densities read

$$(T^0{}_0)_{\text{photon}} = F^{0\mu}F_{0\mu} - \frac{1}{4}g^{\mu\nu}g^{\alpha\beta}F_{\mu\alpha}F_{\nu\beta} = -\frac{1}{2\gamma}\acute{\Pi}^i \gamma_{ij}\acute{\Pi}^j - \frac{1}{N\sqrt{\gamma}}N^j F_{ji}\acute{\Pi}^i - \frac{1}{4}\gamma^{ij}\gamma^{kl}F_{ik}F_{jl},$$

$$(T^0{}_i)_{\text{photon}} = F^{0\alpha}F_{i\alpha} = -\frac{1}{N\sqrt{\gamma}}F_{ij}\acute{\Pi}^j, \qquad (4.22)$$

$$(J^{0i})_{\text{photon}} = \frac{1}{2N\sqrt{\gamma}}\epsilon^{0ijk}F_{jk}.$$

Note that $(J^{0i})_{\text{photon}}$ coinsides with the total magnetic flux densities $J^{0i}$ since the charged matter sector does not contribute. We have specified all the ingredients induced by the two-form background field, which enables us to write down the path-integral formula for the Massieu-Planck functional. With the help of the previous result for the path-integral formula (see the detailed account in Ref. [64]), we obtain the following expression in the axial gauge ($A_3 = 0$):

$$\Psi[\lambda] = \log \int \mathcal{D}\psi \mathcal{D}\bar{\psi}\mathcal{D}A_\mu \prod_{i=1}^3 \mathcal{D}\acute{\Pi}^i \delta(A_3) \det(\nabla_3) \exp\left[\int_0^{\beta_0} d\tau L_{\text{H}}[\varphi, \acute{\Pi}^i]\right], \qquad (4.23)$$

where we introduced the phase-space Lagrangian $L_{\text{H}}$ as

$$L_{\text{H}} \equiv \int d^3x \left[ -\frac{i}{2}\bar{\psi}(\overrightarrow{\partial_\tau} - \overleftarrow{\partial_\tau})\psi + (\acute{\Pi}^i + \epsilon^{0ijk}b_{jk})i\partial_\tau A_i + iA_0\left[\nabla_i \acute{\Pi}^i + \sqrt{-g}j^0 + \frac{1}{2}\nabla_i(\epsilon^{0ijk}b_{jk})\right]\right] - \beta_0^{-1}K$$

$$= \int d^3x e^\sigma u^0 \sqrt{-g}\left[ -\frac{i}{2}\bar{\psi}(\gamma^a \tilde{e}_a{}^\mu \overrightarrow{\tilde{D}}_\mu - \overleftarrow{\tilde{D}}_\mu \gamma^a \tilde{e}_a{}^\mu)\psi - m\bar{\psi}\psi - \frac{1}{4}\gamma^{ij}\gamma^{kl}F_{ik}F_{jl}\right.$$

$$- \frac{1}{2\gamma}\acute{\Pi}^i \gamma_{ij}\acute{\Pi}^j + \frac{1}{\sqrt{-\tilde{g}}}\left[F_{0i} - e^\sigma(u^0 N^j + u^j)F_{ji}\right]\acute{\Pi}^i$$

$$\left. + \frac{1}{2e^\sigma u^0 \sqrt{-g}}\epsilon^{0ijk}F_{0i}b_{jk} + \frac{1}{2e^\sigma u^0 \sqrt{-g}}\epsilon^{0ijk}F_{jk}e^\sigma H_i\right],$$

where we used a decomposed form of the fluid vector $\beta^\mu = \beta u^\mu/\beta_0$ and the reduced magnetic field $\mathcal{H}_\mu = \beta H_\mu$ with a normalized four-velocity $u^\mu$ ($u^\mu u_\mu = -1$) and the ratio of the inverse temperature to the constant reference $e^{\sigma(x)} = \beta(x)/\beta_0$. A new vierbein $\tilde{e}_\mu{}^a$ and its inverse $\tilde{e}_a{}^\mu$ as well as a new covariant derivative $\widetilde{D}_\mu$ will be explained shortly. We also defined the field strength tensor including the imaginary-time direction as $F_{0i} = i(\partial_\tau A_i - \partial_i A_0)$. The use of the axial gauge $A_3 = 0$ leads to the insertion of $\delta(A_3)\det(\nabla_3)$, which is replaced by $\delta(F)\det(\partial F/\partial \alpha)$ in a general gauge satisifying $F = 0$ with the gauge parameter $\alpha$.

Performing the functional integration with respect to the bilinear form of the modified conjugate momentum $\acute{\Pi}^i$, we obtain the path-integral formula for the Massieu-Planck functional:

$$\Psi[\lambda] = \log \int \mathcal{D}\psi \mathcal{D}\bar{\psi} \mathcal{D} A_\mu \delta(F) \det(\partial F/\partial \alpha) \exp\left(\widetilde{\mathcal{S}}_{\text{QED}}[\varphi; \tilde{j}]\right), \tag{4.24}$$

with the imaginary-time action

$$\widetilde{\mathcal{S}}_{\text{QED}} = \int_0^{\beta_0} d\tau d^3 x \sqrt{-\tilde{g}} \left[ -\frac{i}{2}\bar{\psi}(\gamma^a \tilde{e}_a{}^\mu \overrightarrow{\widetilde{D}}_\mu - \overleftarrow{\widetilde{D}}_\mu \gamma^a \tilde{e}_a{}^\mu)\psi - m\bar{\psi}\psi - \frac{1}{4}\tilde{g}^{\mu\nu}\tilde{g}^{\alpha\beta}F_{\mu\alpha}F_{\nu\beta} + \frac{1}{2}\tilde{b}_{\mu\nu}\widetilde{F}^{\mu\nu} \right]. \tag{4.25}$$

This Lagrangian density has the same form as the original one (2.3). This indicates that information on the local thermal equilibrium is thoroughly encoded in the thermally induced background field with the aforementioned vierbein and two-form gauge field $\tilde{b}_{\mu\nu}$ defined by

$$\tilde{e}_0{}^a \equiv e^\sigma u^a, \quad \tilde{e}_i{}^a \equiv e_i{}^a \quad \text{and} \quad \tilde{b}_{0i} \equiv e^\sigma H_i, \quad \tilde{b}_{ij} \equiv b_{ij}. \tag{4.26}$$

Equivalently, the explicit form of the inverse thermal vierbein is given by

$$\tilde{e}_a{}^0 = e_a{}^0 \frac{e^{-\sigma}}{u^0}, \quad \tilde{e}_a{}^i = e_a{}^i - e_a{}^0 \frac{u^i}{u^0}, \tag{4.27}$$

from which one finds the orthogonality of $\tilde{e}_a{}^i$ to the fluid velocity as $u^a \tilde{e}_a{}^i = 0$. In the same way as the original spacetime metric, the emergent thermal metric and its inverse satisfy $\tilde{g}_{\mu\nu} = \eta_{ab} \tilde{e}_\mu{}^a \tilde{e}_\nu{}^b$ and $\tilde{g}^{\mu\nu} = \eta_{ab} \tilde{e}_a{}^\mu \tilde{e}_b{}^\nu$, respectively. Their explicit forms are obtained by replacing the Lapse function $N$ and shift vector $N_i$ in the $(3+1)$ parametrization (4.4) with the following thermal ones:

$$\tilde{N} \equiv N e^\sigma u^0, \quad \tilde{N}_i = e^\sigma u_i, \quad \text{and} \quad \tilde{N}^i = \gamma^{ij}\tilde{N}_j = e^\sigma(u^0 N^i + u^i). \tag{4.28}$$

Here, we used the previously defined dilaton-like parametrization $e^{\sigma(x)} = \beta(x)/\beta_0$. From this parametrization, one can easily extract the measure factor $\sqrt{-\tilde{g}} = \tilde{N}\sqrt{\gamma} = e^\sigma u^0 \sqrt{-g}$. We also introduced a partial derivative in the thermal spacetime as $\tilde{\partial}_\mu = (i\partial_\tau, \partial_i)$ and a covariant derivative for the Dirac field by replacing the spin connection in Eqs. (2.4)-(2.5) with that of thermal spacetime composed of the thermal vierbein (4.26). We note that the constraint (2.11) coming from the local Lorentz symmetry is crucial to obtain the correct expression for the imaginary-time spin connection (see Ref. [64] for the details).

**Symmetries of the Massieu-Planck functional.** Based on the path-integral formula, we demonstrate symmetries of the Massieu-Planck functional. The above result indicates that one needs to perform the path integral in the presence of the emergent background to investigate the transport phenomena taking place in the locally equilibrated QED plasma. Notice that one can summarize information on the background field in the form of the emergent line element $\mathrm{d}\tilde{s}^2$ and two-form gauge connection $\tilde{b}$ in the Kaluza-Klein parametrizations:

$$\mathrm{d}\tilde{s}^2 \equiv \tilde{g}_{\mu\nu}\mathrm{d}\tilde{x}^\mu \otimes \mathrm{d}\tilde{x}^\nu = -\mathrm{e}^{2\sigma}(\mathrm{d}\tilde{t} + a_i\mathrm{d}x^i)^2 + \gamma'_{ij}\mathrm{d}x^i \otimes \mathrm{d}x^j, \tag{4.29}$$

$$\tilde{b} \equiv \frac{1}{2}\tilde{b}_{\mu\nu}\mathrm{d}\tilde{x}^\mu \wedge \mathrm{d}\tilde{x}^\nu = \tilde{b}_{0i}(\mathrm{d}\tilde{t} + a_j\mathrm{d}x^i) \wedge \mathrm{d}x^i + \frac{1}{2}\tilde{b}'_{ij}\mathrm{d}x^i \wedge \mathrm{d}x^j, \tag{4.30}$$

where we introduced $(\mathrm{d}\tilde{t}, \mathrm{d}\tilde{x}) \equiv (-\mathrm{i}\mathrm{d}\tau, \mathrm{d}x)$ and identified the parameters

$$a_i \equiv -\mathrm{e}^{-\sigma}u_i, \quad \gamma'_{ij} = \gamma_{ij} + u_iu_j \quad \text{and} \quad \tilde{b}'_{ij} \equiv b_{ij} - \tilde{b}_{0j}a_i - \tilde{b}_{i0}a_j. \tag{4.31}$$

The Kaluza-Klein parametrization clarifies symmetries of the Massieu-Planck functional since Eqs. (4.29)-(4.30) are invariant under the Kaluza-Klein gauge transformation:

$$\begin{cases} \tilde{t} \to \tilde{t}' = \tilde{t} + \chi(x^i), \\ x^i \to x^{i'} = x^i, \\ a_i(x^i) \to a'_i(x^i) = a_i(x^i) - \partial_i\chi(x^i). \end{cases} \tag{4.32}$$

This Kaluza-Klein gauge symmetry results from the independence of induced background fields $\tilde{j} \equiv \{\tilde{e}_\mu{}^a, \tilde{b}_{\mu\nu}\}$ in the imaginary-time coordinate (note, however, that they are still real-time dependent quantities taking values defined on each equal-time hypersurface of the foliation while its dependence is not explicit in our notations). Investigating the transformation rule induced by the Kaluza-Klein gauge transformation (4.32), one finds that the lower-time and upper-spatial indices are inert while the upper-time and lower-spatial indices transform [41, 63, 64]. For example, the Kaluza-Klein gauge transformation acts on arbitrary vectors $\widetilde{A}_\mu$ and $\tilde{B}^\mu$ in thermal spacetime as

$$\begin{cases} \widetilde{A}_0 \to \widetilde{A}_0, \\ \widetilde{A}_i \to \widetilde{A}_i - \widetilde{A}_0\partial_i\chi, \end{cases} \text{and} \quad \begin{cases} \tilde{B}^0 \to \tilde{B}^0 + \tilde{B}^i\partial_i\chi, \\ \tilde{B}^i \to \tilde{B}^i. \end{cases} \tag{4.33}$$

Noting $a_i = \tilde{g}_{0i}/\tilde{g}_{00}$, one sees the Kaluza-Klein gauge field $a_i$ indeed obeys this transformation rule. Using the Kaluza-Klein gauge field, one can construct the Kaluza-Klein gauge invariant lower-spatial (upper-temporal) component $\widetilde{A}'_i$ ($\tilde{B}^{0'}$) as

$$\widetilde{A}'_i \equiv \widetilde{A}_i - \widetilde{A}_0 a_i, \quad \tilde{B}^{0'} \equiv \tilde{B}^0 + \tilde{B}^i a_i. \tag{4.34}$$

One sees that $\gamma'_{ij}$ and $\tilde{b}'_{ij}$ in Eq. (4.31) are indeed constructed according to this procedure. As will be shown shortly, the Kaluza-Klein gauge invariance restricts possible forms of the Massieu-Planck functional.

In addition to the Kauluza-Klein gauge invariance, the Massieu-Planck functional also enjoys spatial differmorphism

$$x^i \to x^{i'} = f^i(x^i) \tag{4.35}$$

and two-form gauge symmetry applied to the original two-form gauge field $b_{\mu\nu}$ as

$$\tilde{b}_{ij} \to \tilde{b}_{ij} + \partial_i \theta - \partial_j \theta_i. \tag{4.36}$$

We note that only the spatial component of the two-form gauge field, $\tilde{b}_{ij}$, transforms under the original gauge transformation, and $\tilde{b}_{0i}$ is inert under that (recall that $\tilde{b}_{0i}$ does not contains $b_{0i}$). The emergent line element is the same as the previously studied one, and a new ingredient for the QED plasma is the presence of the two-form background connection $\tilde{b}$ induced by the reduced magnetic field $\mathcal{H}_i$.

Furthemore, we also note that there is another restrction coming from discrete symmetry. For instance, the insertion of charge conjugation $\mathsf{C}$ and time reversal $\mathsf{T}$ leads to the following identities:

$$\Psi[\lambda] = \Psi[\Theta_\mathsf{C}^\lambda \lambda], \quad \Psi[\lambda] = \Psi[\Theta_\mathsf{T}^\lambda \lambda], \tag{4.37}$$

where $\Theta_\mathsf{C}^\lambda = \pm 1$ ($\Theta_\mathsf{T}^\lambda = \pm 1$) denotes an eigenvalue of the charge density operators $\hat{c}$ under charge conjugation (time reversal): $\mathsf{C}\,\hat{c}(t,\bm{x})\,\mathsf{C}^{-1} = \Theta_\mathsf{C}^\lambda \hat{c}(t,\bm{x})$ and $\mathsf{T}\,\hat{c}(t,\bm{x})\,\mathsf{T}^{-1} = \Theta_\mathsf{T}^\lambda \hat{c}(t,\bm{x})$. Here, we note that the time reversal transformation is defined as the reflection with respect to the given hypersurface $\Sigma_t$ under consideration, so that the time-argument of the operator $\hat{c}(t,\bm{x})$ is unchanged. We identify $\mathsf{C}$ and $\mathsf{T}$ eigenvalues as

$$\begin{cases} \Theta_\mathsf{C}^{\beta^0} = +1, & \Theta_\mathsf{C}^{\beta^i} = +1, & \Theta_\mathsf{C}^{\mathcal{H}_i} = -1 \\ \Theta_\mathsf{T}^{\beta^0} = +1, & \Theta_\mathsf{T}^{\beta^i} = -1, & \Theta_\mathsf{T}^{\mathcal{H}_i} = -1. \end{cases} \tag{4.38}$$

Equation (4.37) shows the restrictions on possible forms of $\Psi[\lambda]$ from discrete symmetries.

In summary, we have obtained the path-integral formula for the Massieu-Planck functional for the locally equilibrated QED plasma in Eq. (4.24). We can fully capture the effects resulting from inhomogeneous thermodynamic parameters such as the local temperature by the use of the emergent backgrounds with the line element (4.29) and two-form gauge connection (4.30). The background data tells us the symmetries of the Massieu-Planck functional; the Kaluza-Klein gauge symmetry (4.32), spatial diffeormorphism (4.35), and spatial one-form gauge symmetry (4.36). These symmetries together with the discrete-symmetry argument (4.37) constrain possible forms of the Massieu-Planck functional. For instance, one can use $e^\sigma$ and the magnitude of $\tilde{b}_{0i}$ as basic building blocks, which give the local inverse temperature and the magnitude of the reduced magnetic field since they are Kaluza-Klein and one-form gauge invarint quantites. On the other hand, possible forms of the dependences on the Kaluza-Klein gauge field $a_i$ and the spatial component of the two-form gauge field $\tilde{b}_{ij}$ are restricted by the symmetries (4.32)-(4.36). Consequently, dependences on $a_i$ and $\tilde{b}_{ij}$ are allowed only in the form of their field strengths, which are accompanied by, at least, one spatial derivative. We can now find an exhaustive list of the Kaluza-Klein and one-form gauge invariant quantities up to the first-order derivative:

$$O(\partial^0): \begin{cases} \text{scalar}: \ e^\sigma = \beta/\beta_0, \\ \text{spatial vector}: \ \tilde{b}_{0i} = e^\sigma H_i \\ \text{spatial tensor}: \ \gamma^{ij} \ (\text{or } \gamma'_{ij}), \end{cases} \tag{4.39a}$$

$$O(\partial^1): \begin{cases} \text{spatial vector}: \ \nabla_i \sigma, \\ \text{spatial tensor}: \ \nabla_i \tilde{b}_{0j}, \\ \text{spatial tensor}: \ f_{ij} \equiv \partial_i a_j - \partial_j a_i, \\ \text{spatial tensor}: \ h_{ijk} = \nabla_i \tilde{b}'_{jk} + \nabla_k \tilde{b}'_{ij} + \nabla_j \tilde{b}'_{ki}. \end{cases} \quad (4.39b)$$

Since the Massieu-Planck functional enjoys spatial diffeomorphism too, we can construct spatial scalar quantities from these building blocks. In this way, at a given order of the derivative expansion, one can identify the finite number of the invariants compatible with all symmetries. We will explicitly use the symmetry constraints clarified here when we organize a derivative expansion of the Massieu-Planck functional in the subsequent section.

## 4.2 Dissipative part

Let us now look back on the real-time (not imaginary-time) evolution discussed in Sec. 3.3. As we found there, the dissipative part $\langle \delta \hat{T}^\mu{}_\nu \rangle$ and $\langle \delta \hat{J}^{\mu\nu} \rangle$ are associated with the entropy production operator $\hat{\Sigma}[t, t_0; \lambda]$ given in Eq. (3.16). In this section, we focus on this operator and rewrite it in a formal but more suitable form for the derivation of dissipative corrections. According to our matching condition (3.12), dissipative corrections are purely spatial, i.e.,

$$n_\mu \langle \delta \hat{T}^\mu{}_\nu(t, \boldsymbol{x}) \rangle = 0, \quad n_\mu \langle \delta \hat{J}^{\mu\nu}(t, \boldsymbol{x}) \rangle = 0. \quad (4.40)$$

Therefore, the dissipative effects occur in such a way that the conserved charges distributed on the hypersurface $\Sigma_t$ diffuse in the purely spatial direction perpendicular to $n_\mu$. It is thus useful to introduce a projection matrix, which decomposes the curved spacetime index into the time and spatial components as

$$P^\mu_\nu \equiv v^\mu n_\nu + \delta^\mu_\nu, \quad (4.41)$$

where $v^\mu \equiv t^\mu / N$ is the normalized time vector satisfying $v^\mu n_\mu = -1$. Clearly, this matrix has desired orthogonal properties

$$P^\mu_\nu n_\mu = 0 \quad \text{and} \quad P^\mu_\nu v^\nu = 0. \quad (4.42)$$

When one acts the projection matrix on an arbitrary tensor index, a purely spatial index is returned in our coordinate system. Now, by the use of the projection matrix, the derivative operator is decomposed as

$$\nabla_\mu = -\frac{1}{N} n_\mu \nabla_t + \nabla_{\perp \mu} \quad \text{with} \quad \nabla_t = N v^\mu \nabla_\mu \quad \text{and} \quad \nabla_{\perp \mu} \equiv P^\nu_\mu \nabla_\nu, \quad (4.43)$$

Due to the matching condition (4.40), only the spatial derivative $\nabla_{\perp \mu}$ will be shown to appear in the leading-order dissipative corrections. This decomposition allows us to rewrite the entropy production operator (3.16) as

$$\hat{\Sigma}[t, t_0; \lambda] = -\int_{t_0}^t dt' d^3x' \sqrt{-g} \left[ \delta \hat{T}^0{}_\nu \nabla'_t \beta^\nu + \delta \hat{J}^{0\mu} \nabla'_t \mathcal{H}_\mu + \delta \hat{T}^\mu{}_\nu \nabla'_{\perp \mu} \beta^\nu + \delta \hat{J}^{\mu\nu} \nabla'_{\perp \mu} \mathcal{H}_\nu \right], \quad (4.44)$$

where we defined a short-hand notation

$$\nabla_\mu \mathcal{H}_\nu \equiv \nabla_\mu \mathcal{H}_\nu + \frac{1}{2} \beta^\rho H_{\rho\mu\nu}. \tag{4.45}$$

The crucial point here is that the first two terms in Eq. (4.44) contain the time derivatives of parameters $\nabla_t \lambda \equiv \{\nabla_t \beta^\mu, \nabla_t \mathcal{H}_\mu\}$, although it is not obvious at this stage whether those time-derivative terms cause any problem. It will turn out that further manipulation is necessary to eliminate them so that one can obtain the well-defined Green-Kubo formulas for the transport coefficients. In fact, without eliminating those time-derivative terms, they would propagate through the following formulation and finally manifest themselves in the constitutive relations. Then, the current operator contains the contributions from the (linear) hydrodynamic modes, and the gapless hydrodynamic modes contribute to current-current correlators in the resulting Green-Kubo formulas. Consequently, when taking the hydrodynamic limit of correlators, one will suffer from the divergence, and cannot obtain the correct transport coefficients. We will, therefore, appropriately eliminate the time derivatives by the use of the equations of motion (3.10).

Following the procedure developed in Ref. [63], we formally solve the equation of motion (3.10) and express the time derivatives $\nabla_t \lambda$ in terms of the spatial derivatives $\nabla_{\perp\mu}\lambda$. Using the collective (vector-like) notation—the conserved charge density $\hat{c}_a = \{\hat{T}^0{}_\mu, \hat{J}^{0\mu}\}$, conserved current $\hat{\mathcal{J}}_a^\mu = \{\hat{T}^\mu{}_\nu, \hat{J}^{\mu\nu}\}$ satisfying $\nabla_\mu \hat{\mathcal{J}}_a^\mu = \hat{C}_a$, and the parameter derivatives $\nabla_\mu \Lambda^a = \{\nabla_\mu \beta^\nu, \nabla_\mu \mathcal{H}_\mu\}$—we obtain the following formal expression (see Refs. [64, 65] for the details):

$$\nabla'_t \lambda^a(x') = \int d\Sigma_t d\Sigma''_t \chi^{ab}(\boldsymbol{x}', \boldsymbol{x}''; t)(\delta \hat{c}_b(t, \boldsymbol{x}''), \delta \hat{\mathcal{J}}_c^\mu(t, \boldsymbol{x}))_t \nabla_{\perp\mu}\Lambda^c(t, \boldsymbol{x})$$
$$+ \int d\Sigma''_t \chi^{ab}(\boldsymbol{x}', \boldsymbol{x}''; t)\big[\nabla''_\mu \langle \delta \hat{\mathcal{J}}_b^\mu(t, \boldsymbol{x}'') \rangle - \langle \delta \hat{C}_b(t, \boldsymbol{x}'') \rangle\big]. \tag{4.46}$$

We introduced inverse susceptibilities as a second variation of the entropy functional:

$$\chi^{ab}(\boldsymbol{x}, \boldsymbol{x}'; t) = \frac{\delta S[c_a]}{\delta c_a(t, \boldsymbol{x}') \delta c_b(t, \boldsymbol{x}')}, \tag{4.47}$$

and the Kubo-Mori-Bogoliubov (KMB) inner product:

$$(\hat{A}, \hat{B})_t \equiv \int_0^1 d\tau \langle e^{\tau \hat{K}[\lambda_t; t]} \hat{A} e^{-\tau \hat{K}[\lambda_t; t]} \hat{B}^\dagger \rangle_t^{\text{LG}}. \tag{4.48}$$

Equation (4.46) enables us to eliminate the time derivatives in the entropy functional (4.44), which results in the following form:

$$\hat{\Sigma}[t, t_0; \lambda] = -\int_{t_0}^t dt' d^3 x' \sqrt{-g} \Big[\tilde{\delta}\hat{T}^{\mu\nu}\nabla_{\perp\mu}\beta_\nu + \tilde{\delta}\hat{J}^{\mu\nu}\nabla_{\perp\mu}\mathcal{H}_\nu$$
$$+ \delta\hat{\beta}^\nu\Big(\nabla_\mu\langle\tilde{\delta}\hat{T}^\mu{}_\nu\rangle - \frac{1}{2}H_{\nu\alpha\beta}\langle\tilde{\delta}\hat{J}^{\alpha\beta}\rangle\Big) + \delta\hat{\mathcal{H}}_\nu \nabla_\mu\langle\tilde{\delta}\hat{J}^{\mu\nu}\rangle\Big]. \tag{4.49}$$

Here, we defined a projected operator

$$\tilde{\delta}\hat{\mathcal{O}} \equiv (1 - \hat{\mathcal{P}})\delta\hat{\mathcal{O}} \quad \text{with} \quad \hat{\mathcal{P}}\hat{\mathcal{O}} \equiv \int d\Sigma_t \delta\hat{c}_a(x) \frac{\delta}{\delta c_a(x)} \langle\hat{\mathcal{O}}\rangle_t^{\text{LG}}. \tag{4.50}$$

This projection operator $\hat{\mathcal{P}}$ gives the local Gibbs version of Mori's projection operator [79]. We also defined thermodynamically conjugate operators

$$\delta\hat{\beta}^\mu(x) \equiv \int \mathrm{d}\Sigma_t \delta\hat{c}_b(x') \frac{\delta\beta^\mu(x)}{\delta c_b(x')} \quad \text{and} \quad \delta\hat{\mathcal{H}}_\mu(x) \equiv \int \mathrm{d}\Sigma_t \delta\hat{c}_b(x') \frac{\delta\mathcal{H}_\mu(x)}{\delta c_b(x')}. \tag{4.51}$$

Equation (4.49) gives the exact expression for the entropy production operator. As a result, the combination of Eqs. (3.19) and (4.49) gives the exact form of constitutive relations induced by the deviation from the local thermal equilibrium. Note that Eq. (4.49) contains $\langle\tilde{\delta}\hat{T}^\mu_\nu\rangle$ and $\langle\tilde{\delta}\hat{J}^{\mu\nu}\rangle$, so that we need to solve Eqs. (3.19) and (4.49) in a self-consistent manner. However, as discussed in the next section, the terms in the second line of Eq. (4.49) are found to be the second-order derivative corrections with the help of an appropriate power counting scheme. This observation simplifies the procedure to obtain a self-consistent constitutive relation within the first-order derivative expansion.

## 5 Constitutive equations: Derivative expansion

In this section, applying a derivative expansion to the exact results developed in the previous section, we derive the constitutive relations for relativistic magnetohydrodynamic up to the first derivative order.

We first present our power counting scheme. In this paper, we employ the simplest power counting scheme, where all the thermodynamic parameters and background fields are counted as zeroth-order quantities[12]:

$$\lambda = O(\partial^0) \quad \text{and} \quad j = O(\partial^0). \tag{5.1}$$

As a result, one finds that the first line of the entropy production operator (4.49) contains $O(\partial^1)$ contribution, while the second line provides only the higher-order terms (recall that $\langle\tilde{\delta}\hat{T}^\mu_\nu\rangle$ and $\langle\tilde{\delta}\hat{J}^{\mu\nu}\rangle$ contain, at least, one derivative and $H_{\nu\alpha\beta} = O(\partial^1)$ in our counting). Thus, to derive the constitutive relations up to the first derivative, it is sufficient to use the simplified form of the entropy production operator:

$$\hat{\Sigma}[t, t_0; \lambda] = -\int_{t_0}^t \mathrm{d}t' \mathrm{d}^3 x' \sqrt{-g}\Big[\tilde{\delta}\hat{T}^{\mu\nu}\nabla_{\perp\mu}\beta_\nu + \tilde{\delta}\hat{J}^{\mu\nu}\nabla_{\perp\mu}\mathcal{H}_\nu + O(\nabla^2)\Big]. \tag{5.2}$$

### 5.1 Leading order: Ideal magnetohydrodynamics

In the leading-order derivative expansion, we only need to evaluate the expectation value over the local Gibbs distribution [recall that the entropy production (5.24) is accompanied by at least one derivative]. As shown in Eq. (4.13) in a general gauge [or Eq. (4.16) in the hydrostatic gauge], one can extract the local Gibbs average from the Massieu-Planck functional. Moreover, we have already shown the symmetries and building blocks of the Massieu-Planck functional there. In particular, from Eq. (4.39), the symmetries and the number of derivatives allow us to have only two invariant scalars

$$O(\partial^0): \quad e^\sigma = \beta/\beta_0 \quad \text{and} \quad (\tilde{b}_{0i})^2 = \tilde{\gamma}^{\mu\nu}\mathcal{H}_\mu\mathcal{H}_\nu, \tag{5.3}$$

where we used the fact that $\mathcal{H}_0 = 0$ to express $\gamma^{ij}\mathcal{H}_i\mathcal{H}_j$ in the covariant manner. Here, we also introduced an upper-component induced metric and normal vector in thermal spacetime as

$$\begin{cases} \tilde{\gamma}^{\mu\nu} \equiv \tilde{g}^{\mu\nu} + \tilde{n}^\mu\tilde{n}^\nu = \tilde{e}_a{}^\mu\tilde{e}_b{}^\nu(\eta^{ab}+n^a n^b), \\ \tilde{n}^\mu \equiv \tilde{g}^{\mu\nu}\tilde{n}_\mu = \tilde{e}_a{}^\mu n^a, \end{cases} \tag{5.4}$$

whose components can be directly read off from Eq. (4.28). Note that $n^a$ in the second expression is not the thermal vector but the original one with the local Lorentz index. The normalized property of $\tilde{n}_\mu$ in the thermal spacetime, or $\tilde{n}_\mu\tilde{n}_\nu\tilde{g}^{\mu\nu} = -1$, immediately leads to $\tilde{n}_\mu\tilde{\gamma}^{\mu\nu} = 0$. Using these zeroth-order invariants, we identify the leading-order expression of the Massieu-Planck functional:

$$\Psi^{(0)}[\lambda] = \int_0^{\beta_0} d\tau d^3x \sqrt{-\tilde{g}}\, p(\beta,\widetilde{\mathcal{H}}) = \int d^3x \sqrt{\gamma}\, \beta' p(\beta,\widetilde{\mathcal{H}}), \tag{5.5}$$

with $\beta = \sqrt{-g_{\mu\nu}\beta^\mu\beta^\nu}$ and $\widetilde{\mathcal{H}} = \sqrt{\tilde{\gamma}^{\mu\nu}\mathcal{H}_\mu\mathcal{H}_\nu}$. We note again that the metric contracting their indices is different from each other: $g_{\mu\nu}$ for $\beta$, and $\tilde{\gamma}^{\mu\nu}$ for $\widetilde{\mathcal{H}}$. Also, note that the measure factor $\sqrt{-\tilde{g}}$ is necessary to make the volume element $d\tau d^3x\sqrt{-\tilde{g}}$ invariant under the spatial coordinate transformation and the Kaluza-Klein gauge transformation. On the rightmost side, we performed the imaginary-time integration and used $\sqrt{-\tilde{g}} = \tilde{N}\sqrt{\gamma} = \beta'\sqrt{\gamma}/\beta_0$ with $\beta' = -n_\mu\beta^\mu$. With the help of the thermal inverse vierbein and induced metric, we introduce useful variables constructed from $\mathcal{H}_\mu$ as

$$\widetilde{\mathcal{H}}_a \equiv \tilde{e}_a{}^\mu\mathcal{H}_\mu, \quad \widetilde{\mathcal{H}}^\mu \equiv \tilde{\gamma}^{\mu\nu}\mathcal{H}_\nu. \tag{5.6}$$

---

[12] It is possible to employ another power counting scheme. For instance, in order to describe the rapidly rotating QED plasma (but with a small shear flow), one can assign the vortical component as $\nabla_\mu\beta_\nu - \nabla_\nu\beta_\mu = O(\partial^0)$—with small shear $\nabla_\mu\beta_\nu + \nabla_\nu\beta_\mu = O(\partial^1)$—so that the constitutive relation contains the nonperturbative contribution coming from the vorticity. In this case, it is still possible to perform the derivative expansion with respect to $\hat{\Sigma}[t, t_0; \lambda]$ while one needs to take account of a nonperturbative vortical contribution to the Massieu-Planck functional.

Recalling that $u^a\tilde{e}_a{}^i = 0$ as given just after Eq. (4.27) together with $\mathcal{H}_0 = 0$, one finds that the former is orthogonal to the fluid vector: $u^a\widetilde{\mathcal{H}}_a = 0$.

Combining the result in Eq. (5.5) with the variational formula shown in the previous section, we can derive the leading-order constitutive relations that consist *ideal* magnetohydrodynamics[13]. For that purpose, let us then consider the variation with the fixed hypersurface. Useful variational formulas of the various quantities are summarized as

$$\begin{cases} \delta\sqrt{\gamma} = \sqrt{\gamma}e_a{}^\mu\delta e_\mu{}^a, \\ \delta\beta' = -n_\mu\delta\beta^\mu, \\ \delta p = -\dfrac{\partial p}{\partial \beta}(u_\mu\delta\beta^\mu + \beta^\mu u_a\delta e_\mu{}^a) + \dfrac{\partial p}{\partial \widetilde{\mathcal{H}}}\bigl(b^\mu\delta\mathcal{H}_\mu - \widetilde{\mathcal{H}}b^\mu h_a\delta e_\mu{}^a\bigr), \end{cases} \tag{5.7}$$

---

[13]Here, we define our *ideal* magnetohydrodynamics with the constitutive relations at the *non-derivative* (or zeroth) order. The conventional definition of the "ideal" magnetohydrodynamics goes in a different way (see, e.g., Refs. [29–32]) and relies on an illegitimate notation of an "infinite" electrical conductivity that is a dimensionfull quantity and is, moreover, not defined *a priori* in the formulation of hydrodynamics.

where, in addition to $\beta^\mu = \beta u^\mu$, we defined the following decompositions of $\widetilde{\mathcal{H}}_a$ and $\widetilde{\mathcal{H}}^\mu$:

$$\widetilde{\mathcal{H}}_a = \widetilde{\mathcal{H}} h_a, \quad \widetilde{\mathcal{H}}^\mu = \widetilde{\mathcal{H}} b^\mu \quad \text{with} \quad b^\mu \tilde{e}_\mu{}^a h_a = 1. \tag{5.8}$$

For the computation of $\delta p$, we used the variation of the thermal vierbein given by

$$\delta \tilde{e}_\mu{}^a = -\frac{1}{\beta_0 \widetilde{N}} \tilde{n}_\mu e_\nu{}^a \delta \beta^\nu + \delta e_\mu{}^a \tag{5.9}$$

and the orthogonal properties: $u^a h_a = 0$ and $\widetilde{\mathcal{H}}^\mu \tilde{n}_\mu = 0$. Equation (5.9) follows from the fact that the fluid-vector $\beta^\mu$ generally gives the time vector $\tilde{t}^\mu$ in *thermal* spacetime as $\beta^\mu = \beta_0 \tilde{t}^\mu$ (when we use the hydrostatic gauge, it will also match the time vector along the real-time direction). With the help of the above formulas, one can quickly evaluate the variation of the leading-order Massieu-Planck functional (5.5) as

$$\begin{aligned}
\delta \Psi^{(0)} &= \int d^3x \left[ \delta \sqrt{\gamma} \beta' p(\beta, \widetilde{\mathcal{H}}) + \sqrt{\gamma} \delta \beta' p(\beta, \widetilde{\mathcal{H}}) + \sqrt{\gamma} \beta' \delta p(\beta, \widetilde{\mathcal{H}}) \right] \\
&= \int d^3x \beta' \sqrt{\gamma} \bigg[ \left( -\frac{\partial p}{\partial \beta} \beta^\mu u_a + p e_a{}^\mu - \widetilde{\mathcal{H}} \frac{\partial p}{\partial \widetilde{\mathcal{H}}} b^\mu h_a \right) \delta e_\mu{}^a \\
&\quad - \frac{1}{\beta'} \left( \beta' \frac{\partial p}{\partial \beta} u_\mu + p n_\mu \right) \delta \beta^\mu + \frac{\partial p}{\partial \widetilde{\mathcal{H}}} b^\mu \delta \mathcal{H}_\mu \bigg].
\end{aligned} \tag{5.10}$$

Based on the variational formula (4.13), we obtain the constitutive relations of ideal magnetohydrodynamics

$$\langle \hat{T}^\mu{}_a(t,\boldsymbol{x}) \rangle_t^{\text{LG}} = p e_a{}^\mu - \beta \frac{\partial p}{\partial \beta} u^\mu u_a - \widetilde{\mathcal{H}} \frac{\partial p}{\partial \widetilde{\mathcal{H}}} b^\mu h_a, \tag{5.11a}$$

$$\langle \hat{J}^{\mu\nu}(t,\boldsymbol{x}) \rangle_t^{\text{LG}} = \beta \frac{\partial p}{\partial \widetilde{\mathcal{H}}} (u^\mu b^\nu - u^\nu b^\mu), \tag{5.11b}$$

While the first two terms of the energy-momentum tensor (5.11a) are common to the usual relativistic perfect fluid without one-form symmetry, the last term is induced by magnetic one-form symmetry inherent in QED. Although it is not mandatory, one can also extract the conserved charge densities from Eqs. (3.6) and (5.10), which allows us to express $\partial p/\partial \beta$ and $\partial p/\partial \widetilde{\mathcal{H}}$ in terms of conserved charge densities. The constitutive relation (5.11b) describes the local electromagnetic field in the QED plasma flowing with a finite velocity, since $\hat{J}^{0i}$ and $\hat{J}^{ij}$ are the magnetic flux density and electric field, respectively. One then finds only the magnetic flux sits in the fluid rest frame at the ideal order, i.e., as long as the system is in the local thermal equilibrium state.

The new term in Eq. (5.11a) describes a pressure anisotropy induced by the magnetic flux density in the QED plasma. To see this, it is helpful to rewrite the energy-momentum tensor (5.11a) into the following decomposed form:

$$\langle \hat{T}^\mu{}_a(t,\boldsymbol{x}) \rangle_t^{\text{LG}} = e u^\mu u_a + p_\perp (e_a{}^\mu + u^\mu u_a - b^\mu h_a) + p_\parallel b^\mu h_a. \tag{5.12}$$

Here, we defined energy density $e \equiv u_\mu u^a \langle \hat{T}^\mu{}_a(t,\boldsymbol{x}) \rangle_t^{\text{LG}} = -\partial(\beta p)/\partial \beta$ and the pressure components in parallel and perpendicular to the magnetic flux line as

$$p_\perp = p \quad \text{and} \quad p_\parallel = p - \widetilde{H} B, \tag{5.13}$$

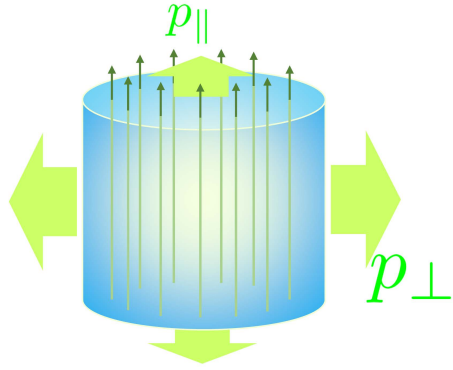

**Figure 1**. Pressure anisotropy in the local thermal equilibrium state with the magnetic flux (green thin arrows). The relative magnitude depends on the the product of the magnetic flux $B$ and magnetic field $H$.

with magnitudes of the magnetic flux density $B \equiv \beta \partial p/\partial \widetilde{\mathcal{H}}$ and effective magnetic field $\widetilde{H} = \beta \widetilde{\mathcal{H}}$. The difference between the parallel and perpendicular components is evident in this expression: $\Delta p \equiv p_\perp - p_\| = \widetilde{H} B$, so that the parallel pressure is effectively reduced (see Fig. 1 for $\widetilde{H} B > 0$). Note again that the matter and magnetic-field contributions are never separated, and $p$ itself contains the contribution of the magnetic flux such as the Maxwell stress and magnetization. Remarkably, $\langle \hat{T}^\mu_{\ a}(t,\boldsymbol{x}) \rangle_t^{\mathrm{LG}} u^a = -e u^\mu$ automatically follows from the orthogonal property $u^a h_a = 0$. Thus, we find that our fluid velocity $u^\mu$ agrees with that in the Landau frame in ideal magnetohydrodynamics.

On the other hand, the energy-momentum tensor in the conventional approach is given by (see Eqs. (14) and (15) of Ref. [33])

$$(T^{\mu\nu})_{\mathrm{conv}} = e_{\mathrm{conv}} u^\mu u^\nu + p_{\mathrm{conv}}(g^{\mu\nu} + u^\mu u^\nu - b^\mu b^\nu) + (p_{\mathrm{conv}} - H_{\mathrm{conv}} B_{\mathrm{conv}}) b^\mu b^\nu, \quad (5.14)$$

where $b^\mu$ is a normalized spatial vector along the magnetic field $B^\mu_{\mathrm{conv}} = B_{\mathrm{conv}} b^\mu$. This expression suggests the following clear identification between our result (5.12) and the conventional one: $\widetilde{H} \leftrightarrow H_{\mathrm{conv}}$, $B \leftrightarrow B_{\mathrm{conv}}$. According to the tensor structures, the energy density $e_{\mathrm{conv}}$ and perpendicular pressure $p_{\mathrm{conv}}$ are also identified with ours. However, since the definition of their variables goes in a different way, we shall briefly summarize the points (see Ref. [33] for detailed accounts). First of all, most of variables in the conventional approach are defined in the additive forms: $e_{\mathrm{conv}} = \varepsilon + \frac{1}{2} B^2_{\mathrm{conv}}$ and $p_{\mathrm{conv}} = p_{\mathrm{matt}} + \frac{1}{2} B^2_{\mathrm{conv}} - M_{\mathrm{conv}} B_{\mathrm{conv}}$. The first and second terms come from the separate contributions of the matter and Maxwell-stress parts. The perpendicular pressure as well as $\varepsilon$ contain a contribution of the medium response, that is, the magnetization potential. Indeed, a part of their energy density $\varepsilon$ is defined with the inclusion of the magnetization potential, i.e, $\varepsilon = \varepsilon_{\mathrm{matt}} - M_{\mathrm{conv}} B_{\mathrm{conv}}$, where $M_{\mathrm{conv}}$ denotes the magnitude of a magnetization vector as mentioned below Eq. (19) of Ref. [33]. The cooperative contributions of the Maxwell stress and magnetization potential lead to the resultant anisotropic pressure $\Delta p_{\mathrm{conv}} = p_{\mathrm{conv}} - (p_{\mathrm{matt}} - \frac{1}{2} B^2_{\mathrm{conv}}) = H_{\mathrm{conv}} B_{\mathrm{conv}}$ with the (in-medium) magnetic field $H_{\mathrm{conv}} = B_{\mathrm{conv}} - M_{\mathrm{conv}}$. All these identifications clarify the breakdown for our energy den-sity and pressure under the assumption of a separation among the matter, Maxwell-stress and magnetization contributions.

## 5.2 Next-to-leading order: Dissipative magnetohydrodynamics

**Derivation of constitutive relations.** Let us proceed to the derivative expansion in the next-to-leading order. One can examine possible derivative corrections from the derivative expansions of the Massieu-Planck functional and the "evolution operator" $\hat{U}[t, t_0; \lambda]$ in Eq. (3.19). However, the charge-neutral QED plasma under consideration is free from the first-order derivative corrections to the Massieu-Planck functional owing to the discrete-symmetry constraint (4.37). Therefore, the local equilibrium parts of constitutive relations are the same as those of ideal magnetohydrodynamics in Eq. (5.11). In the following, we will identify the dissipative contributions to constitutive relations supplemented with the relevant Green-Kubo formula for transport coefficients.

As shown in Eq. (3.16), the "evolution operator" $\hat{U}[t, t_0; \lambda]$ contains $\hat{\Sigma}_\tau[t, t; \lambda]$ in that all the terms have at least one derivative [see Eq. (5.24)]. Therefore, we may write

$$\hat{U}[t, t_0; \lambda] = 1 + \mathrm{T}_\tau \int_0^1 \mathrm{d}\tau \hat{\Sigma}_\tau[t, t; \lambda] + O(\nabla^2). \tag{5.15}$$

Here, we maintain the term up to the first derivative in the current working accuracy, which allows us to use Eq. (5.24) for the entropy production operator. Inserting the above first-order expansion into Eq. (3.19), we have

$$\langle \tilde{\delta} \hat{T}^{\mu\nu}(t, \boldsymbol{x}) \rangle_t = \big( \tilde{\delta} \hat{T}^{\mu\nu}(t, \boldsymbol{x}), \hat{\Sigma}[t, t_0; \lambda] \big)_t + O(\nabla^2), \tag{5.16a}$$

$$\langle \tilde{\delta} \hat{J}^{\mu\nu}(t, \boldsymbol{x}) \rangle_t = \big( \tilde{\delta} \hat{J}^{\mu\nu}(t, \boldsymbol{x}), \hat{\Sigma}[t, t_0; \lambda] \big)_t + O(\nabla^2), \tag{5.16b}$$

where we used the KMB inner product defined in Eq. (4.48).

It is now useful to decompose the projected current operators $\tilde{\delta} \hat{T}^{\mu\nu}$ and $\tilde{\delta} \hat{J}^{\mu\nu}$ by the use of available tensors at hand. The transversality of those dissipative terms to $n_\mu$ greatly restricts possible tensor structures, so that one should utilize the projection matrix $P^\mu_\nu$ defined in (4.41). Besides, we wish to introduce the projection matrix that allows for the decomposition with respect to the direction of the magnetic field. For that purpose, we introduce the lower-component vector $h_\mu \equiv \tilde{e}_\mu{}^a h_a$ from $h_a$, which satisfies $b^\mu h_\mu = 1$ thanks to Eq. (5.8). As was mentioned when we introduced the local Gibbs distribution, the zeroth component of the reduced magnetic field $\mathcal{H}_0$ is set to be zero since the conserved charge $\hat{J}^{0\nu}$ does not have the temporal component $\hat{J}^{00}$. Then, noting $h_\mu = \mathcal{H}_\mu/\widetilde{\mathcal{H}}$ according to its definition, we find that $h_\mu$ is a spatial vector:

$$v^\mu h_\mu = \mathcal{H}_0/\widetilde{\mathcal{H}} = 0. \tag{5.17}$$

Furthermore, using the expression of the original normal vector in terms of the thermal one, $n_\mu = \mathrm{e}^\sigma u^0 \tilde{n}_\mu$, one also finds the upper-index vector $b^\mu$ satisfies $b^\mu n_\mu = 0$. Thus, the vectors $b^\mu$ and $h_\mu$ serve as basic building blocks to prepare the desired projection matrix. Based on this observation, we define another projection matrix

$$\Delta^\mu_\nu \equiv P^\mu_\nu - b^\mu h_\nu = \delta^\mu_\nu + v^\mu n_\nu - b^\mu h_\nu, \tag{5.18}$$

which satisfies the useful properties:

$$\Delta^\mu_\mu = 2, \quad \Delta^\mu_\nu \Delta^\nu_\sigma = \Delta^\mu_\sigma, \quad \Delta^\mu_\nu n_\mu = 0, \quad \Delta^\mu_\nu v^\nu = 0, \quad \Delta^\mu_\nu b^\nu = 0, \quad \Delta^\mu_\nu h_\mu = 0. \quad (5.19)$$

This projection matrix enables us to extract a tensor index pependicular to $n_\mu$, $v^\mu$, $h_\mu$, and $b^\mu$. In short, we find two relevant tensors

$$b^\mu b^\nu \quad \text{and} \quad \Xi^{\mu\nu} \equiv \Delta^\mu_\rho \Delta^\nu_\sigma g^{\rho\sigma}. \quad (5.20)$$

We define $\Xi_{\mu\nu}$ as an "inverse matrix" such that $\Xi^{\mu\rho}\Xi_{\rho\nu} = \Delta^\mu_\nu$.

By using those tensors, we can decompose $\tilde{\delta}\hat{T}^{\mu\nu}$ and $\tilde{\delta}\hat{J}^{\mu\nu}$ as

$$\tilde{\delta}\hat{T}^{\mu\nu} = \tilde{\delta}\hat{p}_\parallel b^\mu b^\nu + \tilde{\delta}\hat{p}_\perp \Xi^{\mu\nu} + 2\tilde{\delta}\hat{\pi}^{(\mu} b^{\nu)} + \tilde{\delta}\hat{\tau}^{\mu\nu}, \quad (5.21a)$$

$$\tilde{\delta}\hat{J}^{\mu\nu} = 2\tilde{\delta}\hat{E}^{[\mu} b^{\nu]} + \tilde{\delta}\hat{D}^{\mu\nu}, \quad (5.21b)$$

where we used $A^{(\mu\nu)} = (A^{\mu\nu} + A^{\nu\mu})/2$ and $A^{[\mu\nu]} = (A^{\mu\nu} - A^{\nu\mu})/2$ to express symmetric and anti-symmetric components. Here, the projected operators are given by

$$\tilde{\delta}\hat{p}_\parallel = \tilde{\delta}\hat{T}^{\mu\nu} h_\mu h_\nu, \quad \tilde{\delta}\hat{p}_\perp = \frac{1}{2}\Xi_{\mu\nu}\tilde{\delta}\hat{T}^{\mu\nu},$$

$$\tilde{\delta}\hat{\pi}^\mu = \Delta^\mu_\rho h_\sigma \tilde{\delta}\hat{T}^{\rho\sigma}, \quad \tilde{\delta}\hat{\tau}^{\mu\nu} = \left(\Delta^\mu_\rho \Delta^\nu_\sigma - \frac{1}{2}\Xi^{\mu\nu}\Xi_{\rho\sigma}\right)\tilde{\delta}\hat{T}^{\rho\sigma}, \quad (5.22)$$

$$\tilde{\delta}\hat{E}^\mu = \Delta^\mu_\rho h_\sigma \tilde{\delta}\hat{J}^{\rho\sigma}, \quad \tilde{\delta}\hat{D}^{\mu\nu} = \Delta^\mu_\rho \Delta^\nu_\sigma \tilde{\delta}\hat{J}^{\rho\sigma}.$$

Note that all the projected vectors and tensors are transverse; that is, $\tilde{\delta}\hat{\pi}^\mu n_\mu = \tilde{\delta}\hat{\pi}^\mu h_\mu = 0$, $\tilde{\delta}\hat{E}^\mu n_\mu = \tilde{\delta}\hat{E}^\mu h_\mu = 0$, $\tilde{\delta}\hat{\tau}^{\mu\nu} n_\nu = \tilde{\delta}\hat{\tau}^{\mu\nu} h_\nu = 0$, and $\tilde{\delta}\hat{D}^{\mu\nu} n_\nu = \tilde{\delta}\hat{D}^{\mu\nu} h_\nu = 0$ are satisfied. The rank-two tensors are symmetric $\tilde{\delta}\hat{\tau}^{\mu\nu} = \tilde{\delta}\hat{\tau}^{\nu\mu}$ and anti-symmetric $\tilde{\delta}\hat{D}^{\mu\nu} = -\tilde{\delta}\hat{D}^{\nu\mu}$, respectively. Likewise, the flow gradient can be decomposed as

$$\nabla_{\perp\mu}\beta_\nu = \theta_\parallel h_\mu h_\nu + \frac{1}{2}\theta_\perp \Xi_{\mu\nu} + 2\Delta^\rho_{(\mu} h_{\nu)} b^\sigma \nabla_{\perp\rho}\beta_\sigma + \left(\Delta^\rho_\mu \Delta^\sigma_\nu - \frac{1}{2}\Xi_{\mu\nu}\Xi^{\rho\sigma}\right)\nabla_{\perp\rho}\beta_\sigma, \quad (5.23)$$

where we defined $\theta_\parallel \equiv b^\sigma b^\sigma \nabla_{\perp\rho}\beta_\sigma$ and $\theta_\perp \equiv \Xi^{\rho\sigma}\nabla_{\perp\rho}\beta_\sigma$. These scalars correspond to the expansion/compression flow in parallel and perpendicular to the magnetic field, respectively. Substituting the decomposed forms (5.21) and (5.23) into Eq. (5.24), we rewrite the entropy production operator as

$$\hat{\Sigma}[t, t_0; \lambda] = -\int_{t_0}^t dt' d^3x' \sqrt{-g} \Big[ \tilde{\delta}\hat{p}_\parallel \theta_\parallel + \tilde{\delta}\hat{p}_\perp \theta_\perp + 2\tilde{\delta}\hat{\pi}^{(\mu} b^{\nu)} \nabla_{\perp\mu}\beta_\nu + \tilde{\delta}\hat{\tau}^{\mu\nu}\nabla_{\perp\mu}\beta_\nu$$
$$+ 2\tilde{\delta}\hat{E}^{[\mu} b^{\nu]} \nabla_{\perp\mu}\mathcal{H}_\nu + \tilde{\delta}\hat{D}^{\mu\nu}\nabla_{\perp\mu}\mathcal{H}_\nu + O(\nabla^2) \Big]. \quad (5.24)$$

We then use this result in Eq. (5.16) to find

$$\int_{t_0}^t dt' d^3x' \sqrt{-g}(\tilde{\delta}\hat{\mathcal{J}}^\mu_a(t, \boldsymbol{x}), \tilde{\delta}\hat{\mathcal{J}}^\nu_b(t', \boldsymbol{x}'))_t \nabla_\nu \lambda^a(t', \boldsymbol{x}')$$
$$= \int_{t_0}^t dt' d^3x' \sqrt{-g}(\tilde{\delta}\hat{\mathcal{J}}^\mu_a(t, \boldsymbol{x}), \tilde{\delta}\hat{\mathcal{J}}^\nu_b(t', \boldsymbol{x}'))_t \nabla_\nu \lambda^a(t, \boldsymbol{x}) + O(\nabla^2). \quad (5.25)$$

Here, we assumed that the integration kernel (or the inner product for the projected current operators) behaves moderately in spacetime. In other words, we assume that $(\tilde{\delta}\hat{\mathcal{J}}_a^\mu(t,\boldsymbol{x}),\tilde{\delta}\hat{\mathcal{J}}_b^\nu(t',\boldsymbol{x}'))_t$ shows a sufficiently rapid decay with respect to spacetime coordinates (e.g. an exponential decay), so that we can apply the Markov approximation. This assumption, expected from the scale separation between hydrodynamic variables (charge densities) and non-hydrodynamic variables (currents), is crucial to obtain the local form of the constitutive relation. It is worth emphasizing that the use of the projection operator (4.50) is essentially important here since it eliminates the linear hydrodynamic mode from the integration kernel. In other words, without the projection operator, we have a contribution from the linear hydrodynamic mode, which causes an ill-mannered behavior of the integration kernel, and thus, prevents us from applying the Markov approximation employed in Eq. (5.25). The subtraction of the linear hydrodynamic mode motivates us to expect the well-mannered behavior described above, but it is fair to say that Eq. (5.25) is still an assumption that we cannot completely verify in this paper[14].

With the help of Eqs. (5.16), (5.24), and (5.25), we finally specify the first-order dissipative corrections to constitutive relations as

$$\langle \tilde{\delta}\hat{T}^{\mu\nu}(t,\boldsymbol{x})\rangle_t = -\frac{1}{\beta}(\zeta_\parallel \theta_\parallel + \zeta_\times \theta_\perp)b^\mu b^\nu - \frac{1}{\beta}(\zeta_\perp \theta_\perp + \zeta'_\times \theta_\parallel)\Xi^{\mu\nu}$$
$$-\frac{2\eta_\parallel}{\beta}\left(b^\mu \Xi^{\nu(\rho}b^{\sigma)} + b^\nu \Xi^{\mu(\rho}b^{\sigma)}\right)\nabla_{\perp\rho}\beta_\sigma - \frac{2\eta_\perp}{\beta}\nabla_\perp^{\langle\mu}\beta^{\nu\rangle}, \quad (5.26a)$$

$$\langle \tilde{\delta}\hat{J}^{\mu\nu}(t,\boldsymbol{x})\rangle_t = -\frac{2\rho_\parallel}{\beta}\left(b^\nu \Xi^{\mu[\rho}b^{\sigma]} - b^\mu \Xi^{\nu[\rho}b^{\sigma]}\right)\nabla_{\perp\rho}\mathcal{H}_\sigma - \frac{2\rho_\perp}{\beta}\Xi^{\mu[\rho}\Xi^{\sigma]\nu}\nabla_{\perp\rho}\mathcal{H}_\sigma, \quad (5.26b)$$

where we introduced the symmetric and traceless projection of a tensor as

$$A^{\langle\mu\nu\rangle} = \frac{1}{2}\left(\Xi^{\mu\rho}\Xi^{\nu\sigma} + \Xi^{\mu\sigma}\Xi^{\nu\rho} - \Xi^{\mu\nu}\Xi^{\rho\sigma}\right)A_{\rho\sigma}. \quad (5.27)$$

Here, we also defined a set of transport coefficients in Eq. (5.26) in the form of the spacetime

---

[14]In fact, the low-dimensional hydrodynamics suffers from the *nonlinear* hydrodynamic fluctuation, which forbids us to derive the local form of the constitutive relation. In this case, we do not have the well-defined transport coefficient in the thermodynamic limit but have the scale-dependent one in the same manner as the running coupling constant in quantum field theory.

integral of the KMB inner product:

$$\zeta_\| = \beta(t,\boldsymbol{x}) \int_{-\infty}^{t} dt' d^3x' \sqrt{-g} (\tilde{\delta}\hat{p}_\|(t,\boldsymbol{x}), \tilde{\delta}\hat{p}_\|(t',\boldsymbol{x}'))_t,$$

$$\zeta_\perp = \beta(t,\boldsymbol{x}) \int_{-\infty}^{t} dt' d^3x' \sqrt{-g} (\tilde{\delta}\hat{p}_\perp(t,\boldsymbol{x}), \tilde{\delta}\hat{p}_\perp(t',\boldsymbol{x}'))_t,$$

$$\zeta_\times = \beta(t,\boldsymbol{x}) \int_{-\infty}^{t} dt' d^3x' \sqrt{-g} (\tilde{\delta}\hat{p}_\|(t,\boldsymbol{x}), \tilde{\delta}\hat{p}_\perp(t',\boldsymbol{x}'))_t,$$

$$\zeta'_\times = \beta(t,\boldsymbol{x}) \int_{-\infty}^{t} dt' d^3x' \sqrt{-g} (\tilde{\delta}\hat{p}_\perp(t,\boldsymbol{x}), \tilde{\delta}\hat{p}_\|(t',\boldsymbol{x}'))_t,$$

$$\eta_\| = \frac{\beta(t,\boldsymbol{x})}{2} \int_{-\infty}^{t} dt' d^3x' \sqrt{-g} (\tilde{\delta}\hat{\pi}^\mu(t,\boldsymbol{x}), \tilde{\delta}\hat{\pi}^\nu(t',\boldsymbol{x}'))_t \Xi_{\mu\nu},$$

$$\eta_\perp = \frac{\beta(t,\boldsymbol{x})}{4} \int_{-\infty}^{t} dt' d^3x' \sqrt{-g} (\tilde{\delta}\hat{\tau}^{\mu\nu}(t,\boldsymbol{x}), \tilde{\delta}\hat{\tau}^{\rho\sigma}(t',\boldsymbol{x}'))_t \Xi_{\mu\rho}\Xi_{\nu\sigma},$$

$$\rho_\| = \frac{\beta(t,\boldsymbol{x})}{2} \int_{-\infty}^{t} dt' d^3x' \sqrt{-g} (\tilde{\delta}\hat{E}^\mu(t,\boldsymbol{x}), \tilde{\delta}\hat{E}^\nu(t',\boldsymbol{x}'))_t \Xi_{\mu\nu},$$

$$\rho_\perp = \frac{\beta(t,\boldsymbol{x})}{2} \int_{-\infty}^{t} dt' d^3x' \sqrt{-g} (\tilde{\delta}\hat{D}^{\mu\nu}(t,\boldsymbol{x}), \tilde{\delta}\hat{D}^{\rho\sigma}(t',\boldsymbol{x}'))_t \Xi_{\mu\rho}\Xi_{\nu\sigma}.$$

(5.28)

These equations give the Green-Kubo formula [67–69] for the transport coefficients of relativistic dissipative magnetohydrodynamics. From their roles in the constitutive relations (5.26), we identify $\zeta$ and $\eta$ as the bulk and shear viscosities, respectively. Recalling that $\epsilon_{0ijk}\hat{J}^{jk}$ represents the electric field while $\epsilon^{0ijk}\nabla_j \mathcal{H}_k \ni \epsilon^{0ijk}\beta^\rho H_{\rho jk} \propto j^i_{\text{bkd}}$ does the background electric current, we can identify $\rho$ with the resistivity. Since the magnetic flux lines break a spatial rotational symmetry, we have two distinct coefficients for the bulk and shear viscosities and the resistivity, i.e., the parallel components $\zeta_\|, \eta_\|, \rho_\|$ and perpendicular ones $\zeta_\perp, \eta_\perp, \rho_\perp$ (see Figs. 2 and 3). Besides, we have two additional cross bulk viscosities $\zeta_\times, \zeta'_\times$, which represent deviation of the pressure in parallel (perpendicular) to the magnetic flux in response to the compression/expansion in the other direction, i.e., the perpendicular (parallel) direction (see Fig. 4). We will show that they should take the same values within the first-order derivative expansion just below.

**Onsager's reciplocal relation.** Since $\zeta_\times$ and $\zeta'_\times$ correspond to the reciprocal processes to each other, one can show that Onsager's reciprocal relation constrains their values [70]. To figure out Onsager's relation for transport coefficients within the first-order derivative expansion, one can reduce $\hat{K}[t;\lambda_t]$ in the Green-Kubo formula (5.28) to a global equilibrium one given by

$$\hat{K}_{\text{eq}}(\beta, \mathcal{H}_i) \equiv \beta\hat{H} - \mathcal{H}_i\hat{\Phi}^i, \quad (5.29)$$

where we introduced a total Hamiltonian $\hat{H}$ and total magnetic flux $\hat{\Phi}^i$ along $x^i$-direction as

$$\hat{H} \equiv -\int d^3x \hat{T}^0{}_0(t,\boldsymbol{x}), \quad \hat{\Phi}^i \equiv -\int d^3x \hat{J}^{0i}(t,\boldsymbol{x}). \quad (5.30)$$

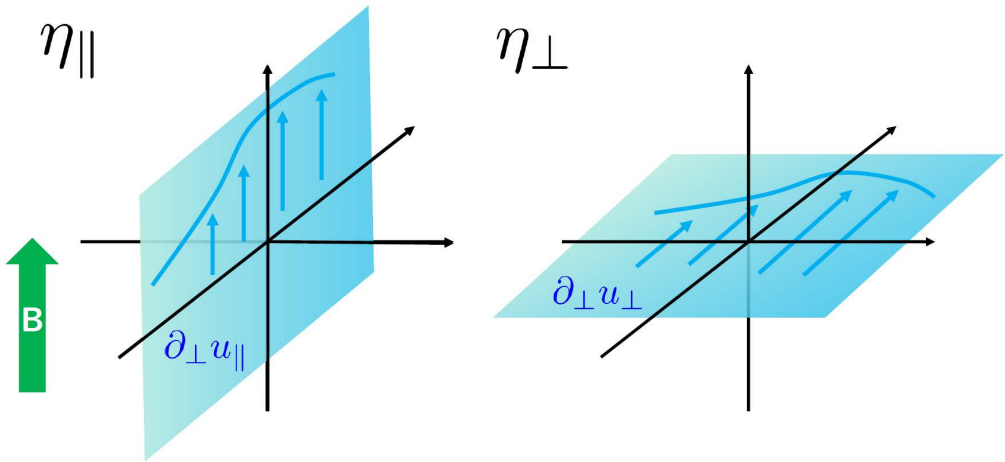

**Figure 2**. Shear deformations in and out of planes with respect to the magnetic flux lines.

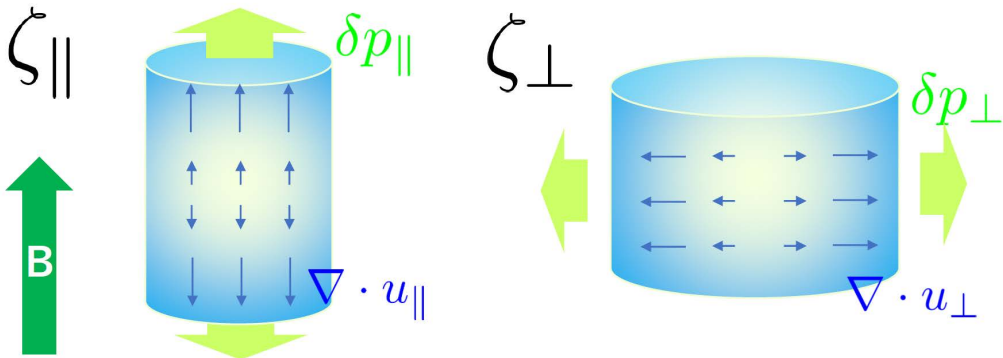

**Figure 3**. Longitudinal and transverse bulk viscosities that are the off-equilibrium responses of the pressure (green arrows) in the direction of the expansion/compression (blue arrows) of the system.

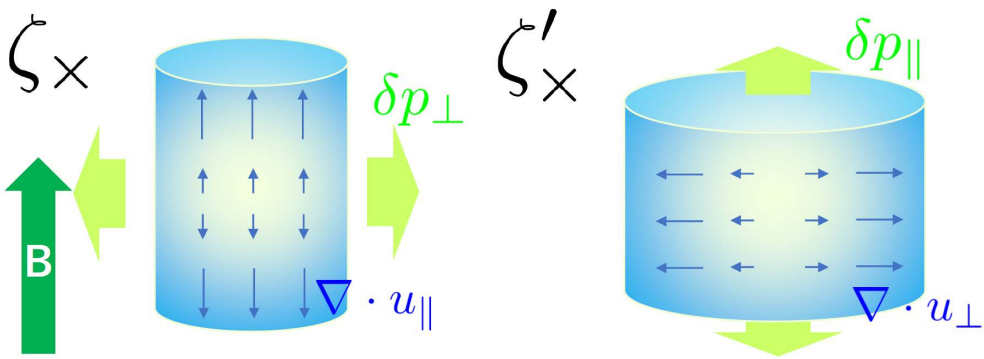

**Figure 4**. Cross bulk viscosities that are the response of the pressure (green arrows) in the orthogonal direction to the expansion/compression (blue arrows) of the system (compare with Fig. 3). Those two cases are reciprocal processes to each other.

We also reduce a spacetime background to the flat spacetime with the Minkowski metric. Then, the Green-Kubo formulas for the cross bulk viscosities read

$$\zeta_\times = \beta \int_0^\infty \mathrm{d}t \int_{-\infty}^\infty \mathrm{d}^3 x \int_0^1 \mathrm{d}\tau \, \mathrm{Tr}\left[ e^{-\hat{K}_{\mathrm{eq}}} e^{\tau \hat{K}_{\mathrm{eq}}} \tilde{\delta} \hat{p}_\parallel(t, \boldsymbol{x}) e^{-\tau \hat{K}_{\mathrm{eq}}} \tilde{\delta} \hat{p}_\perp(0, \boldsymbol{0}) \right],$$
$$\zeta_\times' = \beta \int_0^\infty \mathrm{d}t \int_{-\infty}^\infty \mathrm{d}^3 x \int_0^1 \mathrm{d}\tau \, \mathrm{Tr}\left[ e^{-\hat{K}_{\mathrm{eq}}} e^{\tau \hat{K}_{\mathrm{eq}}} \tilde{\delta} \hat{p}_\perp(t, \boldsymbol{x}) e^{-\tau \hat{K}_{\mathrm{eq}}} \tilde{\delta} \hat{p}_\parallel(0, \boldsymbol{0}) \right]. \quad (5.31)$$

Since the presence of the magnetic flux breaks time-reversal symmetry, i.e., $\mathsf{T} \hat{K}_{\mathrm{eq}}(\beta, \mathcal{H}_i) \mathsf{T}^{-1} = \hat{K}_{\mathrm{eq}}(\beta, -\mathcal{H}_i)$, one may expect Onsager's relation like $\zeta_\times(\beta, \mathcal{H}_i) = \zeta_\times'(\beta, -\mathcal{H}_i)$. Nevertheless, the system is invariant under a successive operation of the charge-conjugation and time-reversal transformation $\Theta \equiv \mathsf{CT}$, i.e.,

$$\Theta \hat{K}_{\mathrm{eq}}(\beta, \mathcal{H}_i) \Theta^{-1} = \hat{K}_{\mathrm{eq}}(\beta, \mathcal{H}_i). \quad (5.32)$$

Taking the time-reversal center as $t = 0$, we also note that $\Theta \tilde{\delta} \hat{p}_\parallel(t, \boldsymbol{x}) \Theta^{-1} = \tilde{\delta} \hat{p}_\parallel(-t, \boldsymbol{x})$ is satisfied. Using those properties, we find

$$\mathrm{Tr}\left[ e^{-\hat{K}_{\mathrm{eq}}} e^{\tau \hat{K}_{\mathrm{eq}}} \tilde{\delta} \hat{p}_\parallel(t, \boldsymbol{x}) e^{-\tau \hat{K}_{\mathrm{eq}}} \tilde{\delta} \hat{p}_\perp(0, \boldsymbol{0}) \right] = \mathrm{Tr}\left[ e^{-\hat{K}_{\mathrm{eq}}} e^{\tau \hat{K}_{\mathrm{eq}}} \tilde{\delta} \hat{p}_\perp(t, -\boldsymbol{x}) e^{-\tau \hat{K}_{\mathrm{eq}}} \tilde{\delta} \hat{p}_\parallel(0, \boldsymbol{0}) \right], \quad (5.33)$$

where we also used the Hermitian properties of the operators. We then reparametrize the spatial coordinate with the opposite sign, which leads to

$$\zeta_\times = \beta \int_0^\infty \mathrm{d}t \int_{-\infty}^\infty \mathrm{d}^3 x \int_0^1 \mathrm{d}\tau \, \mathrm{Tr}\left[ e^{-\hat{K}_{\mathrm{eq}}} e^{\tau \hat{K}_{\mathrm{eq}}} \tilde{\delta} \hat{p}_B(t, -\boldsymbol{x}) e^{-\tau \hat{K}_{\mathrm{eq}}} \tilde{\delta} \hat{p}(0, \boldsymbol{0}) \right] = \zeta_\times'. \quad (5.34)$$

Therefore, we have shown Onsager's relation $\zeta_\times(\beta, \mathcal{H}_i) = \zeta_\times'(\beta, \mathcal{H}_i)$ instead of the naïve one with the flipped magnetic field. As a consequence, we have five (three bulk and two shear) viscosities and two resistivities for the QED plasma preserving the parity and charge-conjugation symmetries. This result from the nonequilibrium statistical operator method provides a basis for that from the phenomenological formulation [35].

**Inequalities for the transport coefficients.** The transport coefficients in Eq. (5.28) satisfy a set of inequalities as a consequence of the positive-definite property of the KMB inner product in the global thermal equilibrium: $(\hat{A}, \hat{A})_{\mathrm{eq}} \geq 0$ (the equality holds for $\hat{A} = 0$). Since all the parallel/perpendicular components of transport coefficients are given by the inner products between the same operators, we immediately find the following constraints:

$$\zeta_{\parallel,\perp} \geq 0, \quad \eta_{\parallel,\perp} \geq 0, \quad \text{and} \quad \rho_{\parallel,\perp} \geq 0. \quad (5.35)$$

On the other hand, the cross viscosity $\zeta_\times (= \zeta_\times')$ is not necessarily a positive-definite quantity since it is given by the inner product between the different operators. However, we can find a constraint on the cross viscosity $\zeta_\times$ by paying attention to the positive-definiteness of the inner product for the following linear combination with an arbitrary real-valued parameter $y$:

$$(\tilde{\delta} \hat{p}_\parallel + y \tilde{\delta} \hat{p}_\perp, \tilde{\delta} \hat{p}_\parallel + y \tilde{\delta} \hat{p}_\perp) \geq 0 \; \Leftrightarrow \; \zeta_\perp y^2 + 2\zeta_\times y + \zeta_\parallel \geq 0. \quad (5.36)$$

Let us first consider the case with $\zeta_\perp > 0$, putting aside the case $\zeta_\perp = 0$. This inequality is satisfied for an arbitrary $y$ if and only if the cross viscosity $\zeta_\times$ satisfies an inequality

$$\zeta_\times^2 - \zeta_\perp \zeta_\| \leq 0. \tag{5.37}$$

When $\zeta_\perp = 0$, the inequality (5.36) is consistent with the inequality for the other bulk viscosity $\zeta_\| \geq 0$ only when $\zeta_\times = 0$. This case can be combined with the inequality (5.37), which implies $\zeta_\times = 0$ when $\zeta_\perp = 0$ (as long as $\zeta_\times$ is a real-valued quantity). The inequality (5.37) implies that $\zeta_\times$ is nonzero only when both of $\zeta_{\|,\perp}$ are finite. As wrap up, we have shown the following set of inequalities

$$\zeta_{\|,\perp} \geq 0, \quad \zeta_\| \zeta_\perp \geq \zeta_\times^2, \quad \eta_{\|,\perp} \geq 0, \quad \text{and} \quad \rho_{\|,\perp} \geq 0. \tag{5.38}$$

If one wishes to find the minimum set of independent inequalities, one may remove either $\zeta_\| \geq 0$ or $\zeta_\perp \geq 0$ from the above list, because one of them, e.g., $\zeta_\| \geq 0$ together with $\zeta_\| \zeta_\perp - \zeta_\times^2 \geq 0$ immediately implies the other inequality $\zeta_\perp \geq 0$, and vice versa.

These inequalities are consistent with those found from the phenomenological analysis in Ref. [35], in which the authors demand the local second law of thermodynamics; that is, the semi-positivity of the local entropy production rate $\nabla_\mu \langle \hat{s}^\mu(t, \bm{x}) \rangle \geq 0$ with the entropy current $\langle \hat{s}^\mu(t, \bm{x}) \rangle$. We shall verify this assumption from our statistical mechanical viewpoint. Since our entropy production operator $\Sigma[t; t_0; \lambda]$ gives a spacetime integral of the local entropy production rate, we can identify $\nabla_\mu \langle \hat{s}^\mu(t, \bm{x}) \rangle$ with the integrand of Eq. (5.24) (see Ref. [63] for a definition of the entropy-current operator). Inserting the first-order constitutive relations (5.26) into the integrand of $\Sigma[t; t_0; \lambda]$ in Eq. (5.24), we obtain the local entropy production rate as

$$\begin{aligned}\nabla_\mu \langle \hat{s}^\mu(t, \bm{x}) \rangle = \frac{1}{\beta} \bigg[ & \begin{pmatrix} \theta_\| & \theta_\perp \end{pmatrix} \begin{pmatrix} \zeta_\| & \zeta_\times \\ \zeta_\times & \zeta_\perp \end{pmatrix} \begin{pmatrix} \theta_\| \\ \theta_\perp \end{pmatrix} \\ & + 2\eta_\| (\nabla_{\perp\mu} \beta_\nu) b^{(\nu} \Xi^{\mu)(\rho} b^{\sigma)} \nabla_{\perp\rho} \beta_\sigma + 2\eta_\perp \nabla_{\perp\langle\mu} \beta_{\nu\rangle} \nabla_\perp^{\langle\mu} \beta^{\nu\rangle} \\ & + 4\rho_\| \nabla_{\perp\mu} \mathcal{H}_\nu b^{[\nu} \Xi^{\mu][\rho} b^{\sigma]} \nabla_{\perp\rho} \mathcal{H}_\sigma + 2\rho_\perp (\nabla_{\perp\mu} \mathcal{H}_\nu) \Xi^{\mu[\rho} \Xi^{\sigma]\nu} \nabla_{\perp\rho} \mathcal{H}_\sigma \bigg]. \end{aligned} \tag{5.39}$$

Note that the right-hand-side has the quadratic form, of which the (matrix) coefficient is given by the set of transport coefficients. Therefore, as a corollary of the set of inequalities (5.38), we can easily verify the semi-positive definiteness of the local entropy production rate—i.e., the local second law of thermodynamics—within the first-order derivative expansion:

$$\nabla_\mu \langle \hat{s}^\mu(t, \bm{x}) \rangle \geq 0. \tag{5.40}$$

While this local positivity constraint is usually postulated as the starting point of the phenomenological formulation of hydrodynamics [28], we have verified it from our statistical mechanical formulation. For reader's convenience, we further compare our viscous coefficients and inequalities to those introduced in Refs. [33, 34]. As discussed in Appendix A, we find a redundancy in the set of inequalities shown in Ref. [34] while we found consistent results in the other parts.

## 6  Summary and Discussions

In this paper, we have investigated a field-theoretical formulation of relativistic magneto-hydrodynamics with the local Gibbs ensemble method (also known as the nonequilibrium statistical operator method). Starting from the QED Lagrangian in the $(3+1)$ dimensional spacetime, we identified the symmetries of the system (including magnetic one-form symmetry) and the local Gibbs density operators relevant for describing the locally equilibrated QED plasma with a dynamical magnetic field. We have provided the exact results for both the local equilibrium and off-equilibrium parts of the constitutive relations, and then performed the derivative expansions for both of them up to the first order. Besides, we have clarified the Green-Kubo formulas, Onsager's reciprocal relation for the cross bulk viscosities $\zeta_\times$, and the inequalities for transport coefficients without relying on the phenomenological assumptions. Here is a short summary of our assumption used in deriving hydrodynamic equations in this paper:

(1) **Special initial condition:** The initial density operator $\hat{\rho}_0$ is assumed to be the local Gibbs distribution, $\hat{\rho}_0 = \hat{\rho}_{\mathrm{LG}}[t_0; \lambda_{t_0}]$ defined in Eqs. (3.2) and (3.3).

(2) **Applicability of derivative expansion 1 (nondissipative part) :** Configurations of thermodynamic parameters are slowly varying functions of spacetime coordinates, and allow us to expand the Massieu-Planck functional as $\Psi[\lambda] = \Psi^{(0)}[\lambda] + O(\nabla^2)$ with Eq. (5.5).

(3) **Applicability of derivative expansion 2 (dissipatve part):** Configurations of thermodynamic parameters are slowly varying functions of spacetime coordinates, and allow us to expand the "evolution operator" $\hat{U}[t, t_0; \lambda]$ with respect to the entropy production $\hat{\Sigma}[t, t_0; \lambda]$ as given in Eq. (5.15).

(4) **Scale separtion and Markov approximation:** The two-point current correlator $(\tilde{\delta}\hat{\mathcal{J}}^\mu_a(t, \boldsymbol{x}), \tilde{\delta}\hat{\mathcal{J}}^\nu_b(t', \boldsymbol{x}'))_t$ is assumed to show a sufficiently rapid decay with respect to spacetime coordinates, so that we can apply the Markov approximation (5.25).

While we have focused on the QED plasma, the same set of hydrodynamic equations may emerge as the universal low-energy effective theory of systems endowed with the same symmetries, i.e., the Poincaré invariance and one-form symmetry.

As concluding remarks, we present some interesting outlooks in order.

**Extension to $n$-group symmetry.** We first note that our framework is applicable to the recently proposed $n$-group symmetry [80]. As a simple example, let us consider the Abelian two-group symmetry equipped with the Poincaré invariance. The conservation laws are generalized as

$$\nabla_\mu T^\mu{}_\nu = \frac{1}{2} J^{\alpha\beta} H_{\nu\alpha\beta} + F^5_{\nu\mu} J^\mu_5, \quad \nabla_\mu J^{\mu\nu} = 0, \quad \nabla_\mu J^\mu_5 = \frac{\hat{\kappa}_{\mathrm{A}}}{2\pi} \varepsilon^{\mu\nu\rho\sigma} F^5_{\mu\nu} J_{\rho\sigma}. \qquad (6.1)$$

Here, we have an additional zero-form symmetry current $J^\mu_5$, which, together with $J^{\mu\nu}$, forms the Abelian two-group symmetry with a certain number $\hat{\kappa}_{\mathrm{A}}$. We also introduced a background field $A_\mu^5$ and its field strength $F_{\mu\nu}^5 \equiv \partial_\mu A_\nu^5 - \partial_\nu A_\mu^5$ coupled with the current $J_5^\mu$. Despite its anomaly-like appearance, the source term in the last equation is actually a consequence of the two-group symmetry (see Ref. [80] in detail). In this setup, one can introduce the conjugate parameter $\nu_5$ to the new charge density $J_5^0$, and the entropy functional takes a generalized form

$$\hat{S}[t;\lambda_t] = -\int \mathrm{d}\Sigma_{t\mu} \left[\hat{T}_\nu^\mu(t,\boldsymbol{x})\beta^\nu(t,\boldsymbol{x}) + \hat{J}^{\mu\nu}(t,\boldsymbol{x})\mathcal{H}_\nu(t,\boldsymbol{x}) + \hat{J}_5^\mu(t,\boldsymbol{x})\nu_5(t,\boldsymbol{x})\right] + \Psi[\lambda_t]. \tag{6.2}$$

Repeating the same procedure presented in this paper, one can, for instance, find the first-order constitutive relation $\langle \delta \hat{J}^{\mu\nu} \rangle$ for the Abelian two-group symmetry by the following replacement in Eq. (5.26b):

$$\nabla_\mu \mathcal{H}_\nu \to \nabla_\mu \mathcal{H}_\nu + \frac{\hat{\kappa}_\mathrm{A} \nu_5}{2\pi} \varepsilon_{\mu\nu\rho\sigma} F^{\rho\sigma,5}. \tag{6.3}$$

This indicates that the background field of zero-form symmetry induces the two-form current. It is notable that the induced two-form current is proportional to the chemical potential $\nu_5$, showing a property similar to the chiral magnetic effect [81–85] (see Refs. [36, 41, 86–91] for hydrodynamic derivations). The complete analysis, including nondissipative transport phenomena, may deserve further studies.

**Spin transport.** There are also several interesting prospects more relevant to familiar physical systems. The first direction is to clarify magnetohydrodynamics with a nonrelativistic component (like the proton in the astrophysical QED plasma) and to investigate roles of their spin degrees of freedom. In the relativistic case, one can interpret Eq. (2.11) as the total angular momentum conservation in the flat spacetime limit. Due to the mutual conversion between the spin and orbital parts of the angular momentum, the spin polarization itself is not a conserved quantity, and does not serve as a strict hydrodynamic variable (see discussions in, e.g., Ref. [92]). Nevertheless, the nonrelativistic (or large-mass) limit leads to an emergent internal SU(2) spin symmetry (like the heavy-quark symmetry [93]), and the spin density for the heavy fermion serves as an emergent hydrodynamic variable. While the interplay between the magnetic field and spin takes place via the Zeeman coupling as the leading correction, which leads to the approximate internal SU(2) symmetry, it is interesting to investigate the coupled dynamics of the nonrelativistic spin density and dynamical magnetic field as in Ref. [94]. Such a hydrodynamic framework will be applied to various systems such as the astrophysical QED plasma composed of (heavy) proton and (light) electron.

**Evaluation of the transport coefficients.** It is also an important direction to investigate the physical properties of the QED plasmas with the help of field-theoretical techniques. While we have clarified the universal forms of the constitutive relations, they contain several physical parameters; the equation of state $p(\beta, \mathcal{H})$ and transport coefficients. As we have presented all the field-theoretical formulas necessary for determining those parameters, one can systematically evaluate them by using, e.g., the finite-temperature perturbation theory [95–100] (see also Refs. [51, 52, 101–104] for the phenomenological treatments of the collisional effects) and/or the strong-coupling methods (see Refs. [105–108] for holographic calculations and Ref. [109] for a recent lattice Monte Carlo simulation).

In evaluation of the Green-Kubo formulas with the finite-temperature perturbation theory [95–100], the magnetic field is often treated as a background field $B_{\rm bkd}$ and is coupled to the charged matter (which gives rise to the Landau quantization in the strong magnetic field). This is in sharp contrast to the current formulation since we fix the value of the dynamical magnetic flux density by using the (reduced) magnetic field, which is not coupled to the charged matter but to the magnetic flux. As a result, effects of the magnetic field on the charged matter are encoded in a different way as the conventional formulation. In many of the above studies, the resummation for the coupling between the charged fermions and magnetic field plays crucial roles. Therefore, it is interesting to push forward with the current "dual" formulation for the evaluation of physical parameters, which could open an avenue for a new resummation scheme. We leave those issues as future works.

## Acknowledgments

The authors thanks Yoshimasa Hidaka, Xu-Guang Huang, Yuta Kikuchi, Shu Lin, Shitaro Sato, and Noriyuki Sogabe for useful discussions. M.H. was supported by the U.S. Department of Energy, Office of Science, Office of Nuclear Physics under Award Number DE-FG0201ER41195. This work was partially supported by the RIKEN iTHEMS Program, in particular iTHEMS STAMP working group. K.H. is supported in part by JSPS KAKENHI under grant No. 20K03948.

## A  Comparison to viscosities in the literature

In this appendix, we present a comparison of the viscous coefficients and associated inequalities obtained in this work to those introduced in the literature [33–35]. As mentioned in Sec. 1, the viscosities for relativistic magnetohydrodynamics have been investigated recently in Refs. [33–35] with the phenomenological formulation. Those authors obtained the same number of independent viscous coefficients but in different tensor bases. Therefore, we shall clarify the correspondences among them (see also Appendix B of Ref. [34]). As a result, we find that our results agree with those obtained in Ref. [35], but there are some discrepancies in the set of inequalities with those of Refs. [33, 34].

The correspondences between the viscous coefficients in Huang, Sedrakian and Rischke (HSR) [33] and Hernandez and Kovtun (HK) [34] are available in Eq. (B.1) of Ref. [34]. Including the viscous coefficients defined in the present paper, the list of correspondences

is expanded as

$$\begin{aligned}
\eta_\perp^{\text{HK}} &= \eta_0^{\text{HSR}} = \frac{1}{2}\eta_\perp\,,\\
\eta_\parallel^{\text{HK}} &= \eta_0^{\text{HSR}} + \eta_2^{\text{HSR}} = \frac{1}{2}\eta_\parallel\,,\\
\eta_1^{\text{HK}} &= -\frac{1}{2}\eta_0^{\text{HSR}} - \frac{3}{8}\eta_1^{\text{HSR}} - \frac{3}{4}\zeta_\perp^{\text{HSR}} = -\frac{1}{2}(\zeta_\perp - \zeta_\times)\,,\\
\eta_2^{\text{HK}} &= \frac{3}{2}\eta_0^{\text{HSR}} + \frac{9}{8}\eta_1^{\text{HSR}} + \frac{3}{4}\zeta_\perp^{\text{HSR}} + \frac{3}{2}\zeta_\parallel^{\text{HSR}} = \frac{1}{2}(\zeta_\parallel - 2\zeta_\times + \zeta_\perp)\,,\\
\zeta_1^{\text{HK}} &= \zeta_\perp^{\text{HSR}} = \frac{1}{3}(2\zeta_\perp + \zeta_\times)\,,\\
\zeta_2^{\text{HK}} &= \zeta_\parallel^{\text{HSR}} - \zeta_\perp^{\text{HSR}} = \frac{1}{3}(\zeta_\parallel + \zeta_\times - 2\zeta_\perp)\,.
\end{aligned} \quad (A.1)$$

The coefficients with the superscripts "HK" and "HSR" are introduced in Refs. [34] and [33], respectively, while those without superscripts are introduced in this paper. We note that the viscous coefficients in the current paper agree with those in Eqs. (3.9), (3.10), (3.13), and (3.14) of Ref. [35].

According to the above correspondences (A.1), it turns out that the quite involved inequalities in Eq. (B.2) of HK [34] can be rewritten with our conventions in drastically simple forms:

$$\eta_\perp \geq 0\,,\quad \eta_\parallel \geq 0\,,\quad \zeta_\perp \geq 0\,,\quad \zeta_\parallel \zeta_\perp - \zeta_\times^2 \geq 0\,,\quad \zeta_\parallel + \zeta_\perp + 2\zeta_\times \geq 0\,. \qquad (A.2)$$

Since one may remove one inequality $\zeta_\parallel \geq 0$ from our list of inequalities (5.38) as mentioned there, an essential difference between the set of inequalities in Eqs. (5.38) and (A.2) is only the existence of the last inequality in Eq. (A.2), i.e., $\zeta_\parallel + \zeta_\perp + 2\zeta_\times \geq 0$. This inequality can be, however, deduced from the other inequalities as follows. According to the inequality $\zeta_\parallel \zeta_\perp \geq \zeta_\times^2$, one can immediately show that $(\zeta_\parallel + \zeta_\perp)^2 - (2\zeta_\times)^2 = (\zeta_\parallel - \zeta_\perp)^2 + 4(\zeta_\parallel \zeta_\perp - \zeta_\times^2) \geq 0$. Since we have $\zeta_{\parallel,\perp} \geq 0$, we can get rid of the square to find the inequality $\zeta_\parallel + \zeta_\perp + 2\zeta_\times \geq 0$, regardless of the sign of $\zeta_\times$. This proof suggests that the fourth inequality in Eq. (A.2), and thus in Eq. (B.2) of HK [34], is a redundant inequality. Therefore, the minimal set of inequalities is given by Eq. (5.38) without either $\zeta_\parallel \geq 0$ or $\zeta_\perp \geq 0$. While all the transport coefficients in Ref. [33] are assumed to be semi-positive definite, this requirement is too strong; that is, not a necessary condition but sufficient one.

Comparison to the viscosities defined in Li and Yee (LY) [98] may be also of reader's interest. In this reference, the authors investigated hydrodynamics in an external *non-dynamical* magnetic field and identified the hydrodynamic modes and relevant viscosities. Since the perpendicular components of momentum charges are not conserved quantities due to the Lorentz force [cf. the discussion below Eq. (1.1)], the number of hydrodynamic modes and of available tensor structures are reduced in the setup of Ref. [98]. Although we should stress the difference between the dynamical and nondynamical magnetic fields, the correspondences to the three viscous coefficients in Eq. (A5) of Ref. [98] are found to be

$$\eta^{\text{LY}} = \eta_\parallel,\quad \zeta^{\text{LY}} = \zeta_\parallel,\quad \zeta'^{\text{LY}} = \zeta_\times, \qquad (A.3)$$

where the superscripts LY denote the viscosities introduced in Ref. [98]. It is clear that all of them quantify the responses to flow perturbations along the external magnetic field. The cross viscosity $\zeta_\times = \zeta^{\text{LY}}$, however, contributes only to the transverse pressure, i.e., the transverse components of the energy-momentum tensor that are not conserved currents in the setup of Ref. [98]. Therefore, the cross viscosity may not be allowed in the strict hydrodynamic limit. Nevertheless, there could be a potential relevance of the cross viscosity if one considers the transverse momentum as a quasi-hydrodynamic mode. An entropy-current analysis with a decent order-counting scheme for this specific setup is necessary for clarifying this point. From our point of view, if the cross viscosity contributes to the local entropy production at the first order, the cross bulk viscosity $\zeta_\times = \zeta^{\text{LY}}$ should vanish according to the inequality (5.37) with $\zeta_\perp = 0$ for the semi-positive definite entropy production.